\documentclass[twocolumn,prl,floatfix,longbibliography,showpacs,superscriptaddress]{revtex4-2}

\usepackage[caption=false]{subfig} %
\usepackage{ragged2e} %
\DeclareCaptionJustification{justified}{\justifying}

\usepackage{bm} 
\usepackage{subfig}
\usepackage{amsmath}
\usepackage{amsfonts}
\usepackage{amsbsy}
\usepackage{xcolor}
\usepackage{graphicx}
\usepackage{epsfig}
\usepackage{hyperref}
\usepackage{mathtools}
\hypersetup{
    unicode=true, %
    plainpages=false,
    colorlinks=true,%
    linkcolor=blue,%
    citecolor=blue,%
    filecolor=blue,%
    urlcolor=blue%
}
\usepackage{amsmath}
\usepackage{amsfonts}
\usepackage{amsbsy}
\usepackage{xcolor}
\usepackage{graphicx}
\usepackage{epsfig}

\newcommand{\beq}{\begin{equation}}
\newcommand{\eeq}{\end{equation}}
\newcommand{\bea}{\begin{eqnarray}}
\newcommand{\eea}{\end{eqnarray}}
\newcommand{\BEQAL}{\begin{align}}
\newcommand{\EEQAL}{\end{align}}

\makeatletter
\renewcommand{\fnum@figure}{\textbf{Fig.} \thefigure}
\makeatother

\renewcommand{\thefigure}{\textbf{\arabic{figure}}}

\makeatletter
\def\@hangfrom@section#1#2#3{\@hangfrom{#1#2}#3}%
\def\@hangfroms@section#1#2{#1#2}%
\makeatother

\begin{document}
\let\arxivAddToContents\addtocontents
\renewcommand{\addtocontents}[2]{}

\title{Semiclassical Phase-Space Dynamics of Emitter Ensembles with Local Dissipation}
\author{L. Ruks}
\affiliation{Basic Research Laboratories \& NTT Research Center for Theoretical Quantum Information, NTT, Inc., 3-1 Morinosato Wakamiya, Atsugi, Kanagawa, 243-0198, Japan}

\begin{abstract}
We establish that permutation-invariant emitter ensembles with local dissipation admit a semiclassical description on the four-dimensional phase space of a variable-length collective spin, within a perturbative truncated Wigner approximation (TWA). The TWA captures nonlinear dynamics and nonclassical signatures without prescribing closures of correlations. We cast a class of spatially extended emitter systems as interacting ensembles tractable in the TWA to show how quantum fluctuations reshape directional cooperative emission beyond mean-field predictions in a pumped chain containing tens of thousands of emitters, using hundreds of phase-space variables. Our results offer a unified description of collective dynamics beyond the classical limit in emitter ensembles.
 
\end{abstract}
\date{October 7, 2026}
\maketitle

\textbf{Introduction}---Ensembles of emitters are a fundamental element for understanding and engineering open quantum systems. In atomic~\cite{Bohnet2012,Baumann2010,herskind09,garttner17} and solid-state~\cite{Scheibner2007,zhu11,lei23,quach22,kakuyanagi16} realizations, collective interactions generally coexist with local dissipation. While this dissipation must often be mitigated, the interplay with collective dynamics can enable novel dissipative phases of matter~\cite{wang25,keeling17,Lee14,xu14,lee12} and device functionalities enhanced by cooperativity~\cite{meiser09,ostermann13,tucker20}. \\
\indent As mesoscopic emitter ensembles come increasingly into focus, exact treatments~\cite{chase2008,shammah2018,Sarkar1987a,Xu2013,keeling17,wang25,gegg17} exploiting permutation invariance are typically constrained to modest system sizes. This limitation has motivated methods~\cite{weimer21,plankensteiner22,verstraete04,zwolak04,Zhang25,singh22,huber22,merkel20,carmichael2008,huber21} for reducing many-body dynamics. Often, tractable descriptions are obtained by prescribing correlations expected near a classical thermodynamic limit. However, nonclassical correlations and macroscopic, non-Gaussian fluctuations, especially in dynamics, continue to pose a challenge even in large ensembles~\cite{huber21,stitely23,hannukainen18,huybrechts20,Zens2019,fasser25}. Against this backdrop, perturbative semiclassical methods have attracted renewed attention~\cite{mink22,kolmer2024,barberena2025,hosseinabadi25,noel2026} as approaches to systematically capture such corrections. \\
\indent The truncated Wigner approximation (TWA) constitutes the leading order of the Wigner--Moyal semiclassical expansion~\cite{moyal49,rundle21,polkovnikov10}. Here, quantum fluctuations are perturbatively incorporated into an ensemble of trajectories on the system's classical phase space, with initial fluctuations represented by random sampling and dissipation inducing stochastic evolution. For bosons, the TWA has a canonical formulation~\cite{steel98,filipowicz86,blakie08}. For permutation-invariant (PI) emitter ensembles, the two-dimensional Bloch sphere has long underpinned the TWA for purely collective dynamics~\cite{haake87,varilly89,altland09}. Yet this phase space cannot accommodate changes in total spin and associated fluctuations induced by local dissipation. To our knowledge, an analogous TWA formulated for PI ensembles that includes local dissipation has remained absent, constraining investigations of their nonclassical dynamics. \\
\begin{figure}[h!]
    \centering
    \includegraphics[width=0.99\linewidth]{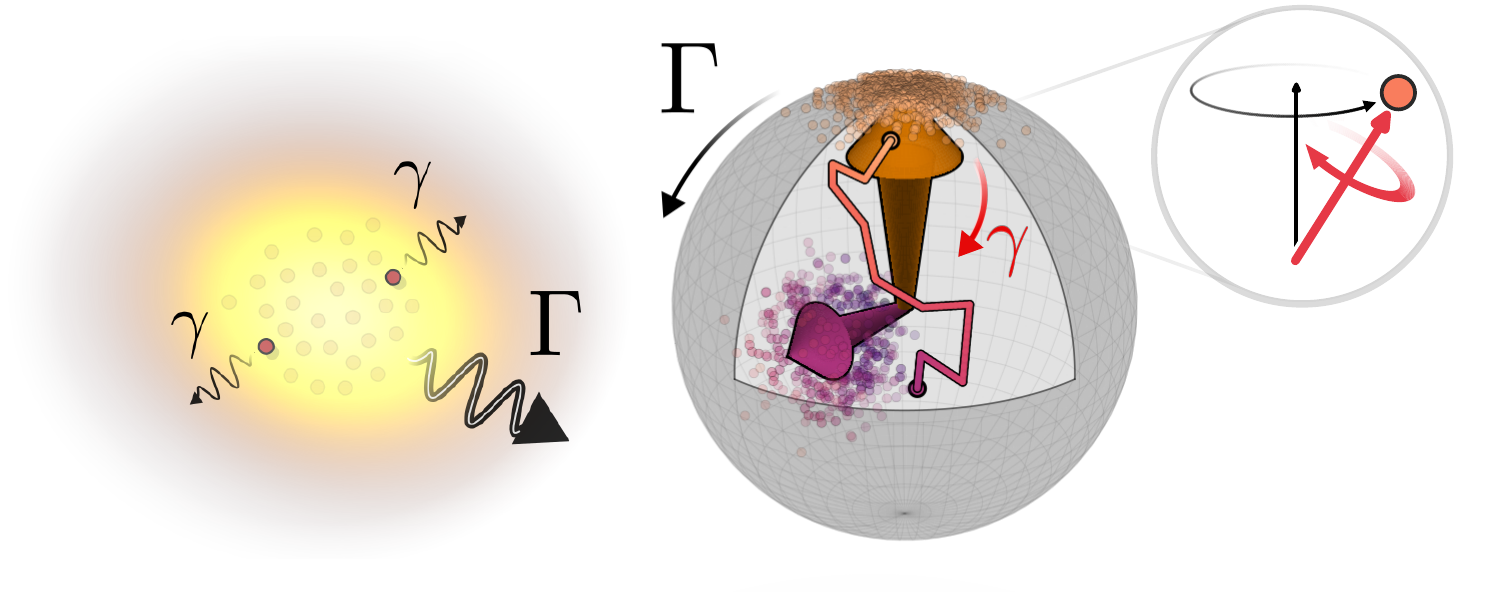}
    \caption{(Left) An ensemble of emitters couples uniformly to a common field mode. The ensemble emits collectively with rate $\Gamma$ and is subject to homogeneous local dissipation with rate $\gamma$. (Right) In the TWA, the ensemble's collective spin (large arrow) is sampled by phase-space variables (dots). In phase space, spin inversion and its conjugate angle (inset, black) suffice to capture (stochastic) evolution on the Bloch-sphere surface under collective processes. A dynamical total-spin variable and its conjugate angle (red) capture changes in total spin under local dissipation, which drives trajectories into the sphere's interior.}
    \label{fig:fig1}
\end{figure}
\indent In this work, we establish a semiclassical limit for the dynamics of a PI emitter ensemble with local dissipation, through a Wigner–Moyal expansion on the four-dimensional phase space of a single variable-length collective spin~\cite{klimov2008,tomatani2015}. 
In the resulting TWA, two phase-space variables---a total-spin variable and its conjugate angle---extend the Bloch sphere to capture changes in total spin, and all dissipation is represented by a few stochastic processes (Fig.~\ref{fig:fig1}), with strictly nonnegative diffusion. In central models including bad-cavity superradiance, the TWA accurately captures nonlinear dynamics in ensembles with as few as tens of emitters—including nonclassical squeezing and subradiance—when prescribed low-order closures of correlations lose accuracy. By expressing a class of spatially extended media in terms of interacting PI ensembles, we obtain a TWA in which the number of phase-space variables scales with the number of distinct ensembles, rather than emitters. In a one-dimensional chain with tens of thousands of emitters, quantum fluctuations shape directional selection of cooperative emission under uniform pumping and, in contrast to mean-field predictions, sustain backward emission under forward coherent driving.

\textbf{Obtaining the TWA for PI ensembles}---We consider a PI ensemble formed from $N$ two-level emitters indexed by $n$. Coupling to common modes induces coherent evolution under a Hamiltonian $\hat H$ and collective dissipation via jumps $\hat L^{q}$ built from collective spin operators
$\hat{L}^{0}=\hat{J}^{z} = \frac{1}{2}\sum_{n}\hat{\sigma}_{n}^{z}, \ 
\hat{L}^{\pm 1}= \hat{J}^{\pm} = \sum_{n}\hat{\sigma}_{n}^{\pm}$,
where $\hat{\sigma}^{0}_{n} \equiv \hat{\sigma}^{z}_{n}$ and $\hat{\sigma}^{\pm 1}_{n} \equiv \hat{\sigma}^{\pm}_{n}$ are the Pauli-$z$ and ladder operators of emitter $n$. Including identical, independent local dissipation, the system evolves under the Born–Markov master equation ($\hbar = 1$),
\begin{equation}
    \label{eq:master-equation-explicit}
     \dot{\hat{\rho}} = -i\left[\hat{H},\hat{\rho}\right] + \sum_{q}{\frac{\Gamma^{q}}{2}}\mathcal{D}\left[\hat{L}^{q}\right]\hat{\rho} + \sum_{q,n}{\frac{\gamma^{q}}{2}}\mathcal{D}\left[\hat{\sigma}_{n}^{q}\right]\hat{\rho}.
\end{equation}
Here, values $q=-1,0,1$ label decay, dephasing, and pumping, respectively, and $\mathcal{D}[\hat{X}]\hat{\rho} = 2\hat{X}\hat{\rho}\hat{X}^{\dagger} - \{\hat{X}^{\dagger}\hat{X},\hat{\rho}\}$.

To obtain a semiclassical approximation of the PI dynamics in Eq.~\eqref{eq:master-equation-explicit}, we use the equivalent angular-momentum representation~\cite{shammah2018,Baragiola2010,Xu2013,supplement}. Here, any PI density matrix can be expressed as
\begin{equation}
\hat{\rho} = \sum_{JMM'}\rho_{JMM'}|JM\rangle \langle JM'|.
\label{eq:density}
\end{equation}
In this representation, $|JM\rangle$ denotes collective kets labeled by eigenvalues $J(J+1)$ and $M$ of total spin squared $\hat J^{2}$ and inversion $\hat J^{z}$, respectively. Elements $|JM\rangle \langle JM'|$ correspond to degeneracy-averaged Dicke-state outer products in the full Hilbert space. $\hat{\rho}$ is then block diagonal in $J$, and collective operators $\hat{J}^{q}$ act as rotations in each block. Local dissipation induces population transfer between distinct sectors $J \leq N/2$ and can be expressed~\cite{kolmer2024} using effective jumps $\hat{l}^{jq}$ in the representation Eq.~\eqref{eq:density}:
\begin{equation}
\label{eq:master-equation-symmetric-basis}
\sum_{n}\mathcal{D}\left[\hat{\sigma}_{n}^{q}\right]\hat{\rho} = \sum_{j}\mathcal{D}\left[\hat{l}^{jq}\right]\hat{\rho}.
\end{equation}
For each $q$, the three jump operators $\hat{l}^{jq}$ ($j = -1,0,1$) couple both inversion $M\to M+q$ and total-spin number $J\to J+j$ acting on elements $|JM\rangle$, completing the angular-momentum representation of Eq.~\eqref{eq:master-equation-explicit}.

The representation, Eq.~\eqref{eq:density}, allows us to employ the Stratonovich--Weyl (SW) correspondence developed for variable-spin systems~\cite{klimov2008} in the PI dynamics, Eq.~\eqref{eq:master-equation-explicit}. For the SW kernel $\hat{\Delta},$ each operator $\hat O$ is assigned a Weyl symbol $\mathcal O(\bm{\mathcal Z})=\mathrm{Tr}[\hat O\,\hat\Delta(\bm{\mathcal Z})]$ and the density matrix is assigned a Wigner function $W(\bm{\mathcal Z})=\mathrm{Tr}[\hat\rho\,\hat\Delta(\bm{\mathcal Z})]$ of classical variables $\bm{\mathcal{Z}}$. For a collective spin, the kernel takes the covariant form~\cite{tilma2016,klimov17,brif99} $\hat\Delta(\bm{\mathcal Z})=\hat U(\bm{\Omega})\hat\Pi_{\mathcal J}\hat U^\dagger(\bm{\Omega})$, for the collective rotation $\hat{U}(\bm{\Omega})$ by Euler angles $\bm{\Omega}=(\phi,\theta,\psi)$. Here, $\hat{U}$ acts on a generalized parity operator $\hat\Pi_{\mathcal J}$ that incorporates couplings between elements in Eq.~\eqref{eq:density} with distinct $J$. The Weyl symbol ${\mathcal J}({\mathcal J}+1)$ of $\hat J^2$ identifies the half-integer $\mathcal J =0,1/2,\ldots$ as a classical total-spin variable analogous to $J$. To clarify the role of $\psi$, we calculate Weyl symbols of local (collective) operators $\hat l^{jq}$ ($\hat L^q$):
\begin{equation}
    \ell^{jq} = \chi_{jq}(\mathcal{J})\, D^{1*}_{qj}(\phi,\theta,\psi), \quad 
    \mathcal{L}^{q} = \chi_{q}(\mathcal{J})\, D^{1*}_{q0}(\phi,\theta).
    \label{eq:weyl-symbol}
\end{equation}
Here, $D^{1}_{qj}$ is the spin-1 Wigner-$D$ matrix element and $\chi$ represents jump amplitudes~\cite{supplement}. Crucially, dependence on $\psi$ arises only in matrix elements with $j \neq 0$, corresponding to jumps $\hat{l}^{jq}$ that change total spin. The spherical angles $(\phi,\theta)$ on the Bloch sphere capture collective dynamics for $\gamma^{q} = 0$, while variables $(\psi,\mathcal{J})$ are required to describe total-spin mixing under local dissipation~\cite{barberena2025}. As $\psi$ is absent from Weyl symbols of collective observables, it serves as an auxiliary variable in practice~\cite{davidson15}. \\
\indent With a complete quantum--classical mapping specified by the representation in Eq.~\eqref{eq:density}, we apply a semiclassical Wigner--Moyal expansion of Eq.~\eqref{eq:master-equation-explicit}. Treating $\mathcal{J}\gg 1$ as continuous yields the approximation~\cite{tomatani2015},
\begin{equation}
\label{eq:semiclassical-commutator}
\mathrm{Tr}\!\left[\hat O\hat O'\hat\Delta\right]\approx \mathcal O\,\mathcal O' + \frac{i}{2}\{\mathcal O,\mathcal O'\},
\end{equation}
where the Poisson bracket $\{\bullet,\bullet\}$ corresponds to a linear rigid rotor for the four-dimensional phase-space coordinates $\bm{\mathcal Z}=(\phi,\theta,\psi,\mathcal J)$ (Appendix).
When total spin is macroscopic and Weyl symbols vary slowly on the effective Planck scale, relative corrections $O(\mathcal{J}^{-1})$ are typically of order $N^{-1}$, providing a semiclassical control parameter. Applying the SW correspondence to Eq.~\eqref{eq:master-equation-explicit} and using Eq.~\eqref{eq:semiclassical-commutator} defines a Fokker–Planck equation for $W$. The diffusion is nonnegative by construction~\cite{dubois2021,supplement}, enabling a stochastic unraveling and yielding the semiclassical Langevin equation for arbitrary PI ensembles:
\begin{equation}
\begin{aligned}
\frac{d\bm{\mathcal{Z}}}{dt} &= \{ \bm{\mathcal{Z}},\mathcal{H}\}
+ \mathrm{Re}\!\left[\sum_{k} \left\{\bm{\mathcal{Z}},\sqrt{r_{k}}\mathcal{L}_{k}\right\} \circ
\Bigl(\eta_{k} + i\sqrt{r_{k}}\mathcal{L}_{k}^{*} \Bigr)\right].
\end{aligned}
\label{eq:SDE}
\end{equation}
Here, $\eta_k$ are independent complex white noise processes, $\langle \eta_{k}(t) \eta^{*}_{k'}(t')\rangle/2 = \delta_{kk'}\delta(t-t')$, with $\circ$ denoting Stratonovich multiplication. $(\mathcal{L}_k,r_k)$ are the Weyl symbols and corresponding rates of jump operators, $\mathcal{L}_{k} = \ell^{jq}, \mathcal{L}^{q}, r_{k} = \gamma^{q},\Gamma^{q}$, for the dissipation channel index $k$, while $\mathcal{H}$ is the Weyl symbol of the Hamiltonian.

\textbf{Applying the TWA}---To estimate expectations of observables for the dynamics in Eq.~\eqref{eq:master-equation-explicit} in this semiclassical description, we average Weyl symbols over $N_{\mathrm{traj}}$ trajectories
$\bm{\mathcal Z}^{(i)}(t)$ that approximately sample $W$ at time $t$,
\begin{equation}
\langle \hat O(t)\rangle
\approx  \frac{1}{N_{\mathrm{traj}}}\sum_{i=1}^{N_{\rm traj}}
\mathcal O(\bm{\mathcal Z}^{(i)}(t)),
\label{eq:expectations}
\end{equation}
with the four phase-space variables initialized to sample the approximate Wigner function of a fully polarized state~\cite{supplement} and evolved using Eq.~\eqref{eq:SDE}. In the following, $N_{\mathrm{traj}} = 10^3\text{--}10^6$ typically renders sampling errors of order $1/\sqrt{N_{\mathrm{traj}}}$ negligible, with smaller errors at larger $N$. \\
\textbf{Validating the TWA}---To validate this semiclassical description, we have benchmarked the TWA in settings ranging from the purely collective limit, where it recovers existing formulations~\cite{altland09,dubois2021}, to driven-dissipative and strongly nonlinear spin dynamics~\cite{supplement}. Here, we illustrate its key features in the model of a bad-cavity superradiant laser~\cite{meiser09,Bohnet2012}, in which collective decay $\Gamma$ (for $q=-1$) competes with local pumping $\gamma$ ($q=1$), and $\hat{H} = 0$. These dynamics admit exact simulations~\cite{zhang18}, allowing us to assess the TWA for thousands of emitters. \\
\begin{figure}
    \centering
    \includegraphics[width=0.95\linewidth]{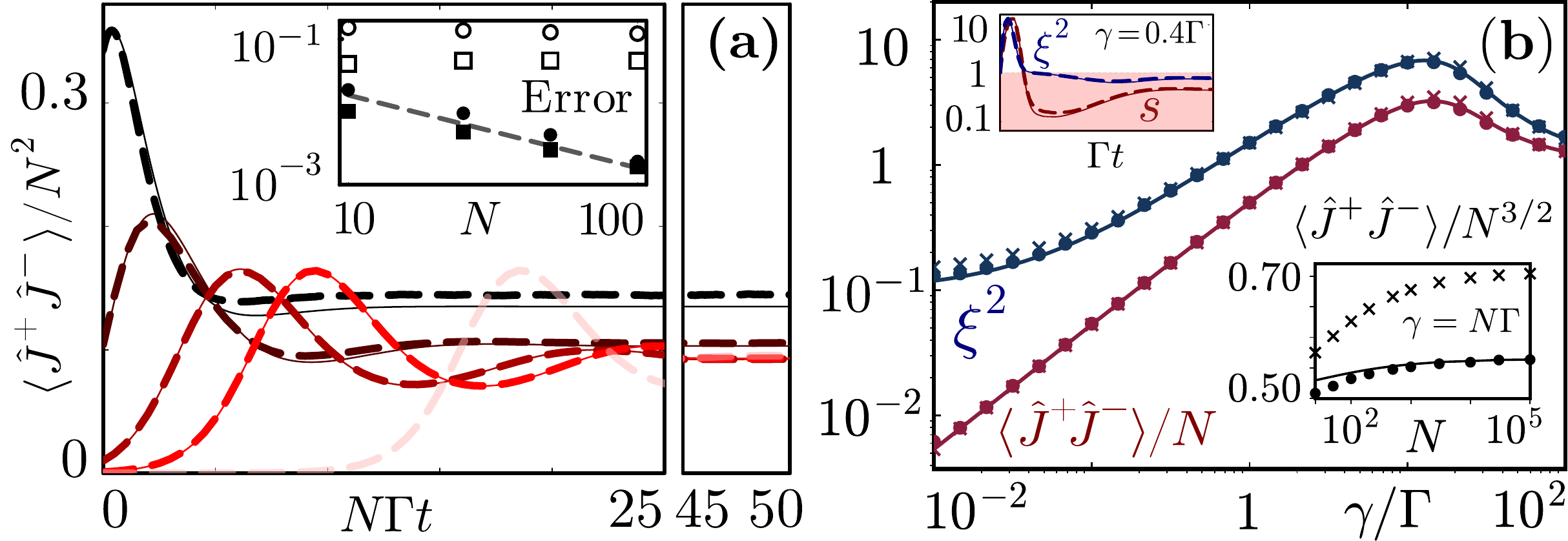}
    
    \caption{Dynamics of a fully inverted PI ensemble under local pumping and collective decay. (a) Evolution of normalized emission rate $\langle \hat{J}^{+}\hat{J}^{-}\rangle/N^2$ for $N=3, 10, 10^2, 10^3,$ and $10^6$ (dark to light), for scaled time $N\Gamma t$ (axis break as indicated) and pump $\gamma = 0.25N\Gamma$. Solid curves show exact dynamics and dashed curves show the TWA. For $N=10^6$, exact simulation is not feasible here. Inset: maximum absolute error between TWA and exact dynamics for the normalized emission rate (solid squares) and normalized inversion $\langle \hat{J}^{z}\rangle/N$ (solid circles) over the full evolution, compared to a $1/N$ reference (dashed). Hollow points correspond to the mean-field prediction. (b) Steady-state squeezing $\xi^2$ and emission rate $\langle \hat{J}^{+}\hat{J}^{-}\rangle/N$ versus pump for TWA (dots), second-order cumulant expansion (crosses), and exact results (solid line), with $N=25$. Lower inset shows the change with $N$ at pump $\gamma = N\Gamma$, and upper inset gives evolution of squeezing and subradiance parameter $s$ up to $\Gamma t = 4$, for $\gamma = 0.4\Gamma$.}

    \label{fig:fig2}
\end{figure}
\indent In Fig.~\ref{fig:fig2}(a), we present the evolution of normalized emission rate $\langle\hat J^+\hat J^-\rangle/N^2$ for a fully inverted ensemble as $N$ is increased from three up to one million. The pump $\gamma = 0.25N\Gamma$ is chosen to maintain competition with decay. Already for $N = 3$, the TWA captures, with less than 10\% deviation, the transient superradiant pulse, subsequent relaxation, and the collective steady-state emission. For $N \simeq 10$, predictions are visibly close to exact dynamics. To compare with a classical prediction [Fig.~\ref{fig:fig2}(a), inset], we keep only the leading order in $\mathcal J^{-1}$ in the deterministic thermodynamic limit of Eq.~\eqref{eq:SDE} to recover the mean-field equations of motion for collective-spin Weyl symbols $\mathcal J^\alpha$. This limit, however, retains only the initial fluctuations and yields larger errors that, unlike those of the TWA, do not decrease with ensemble sizes over the range shown. \\
\indent To show qualitatively nonclassical dynamics in the TWA, Fig.~\ref{fig:fig2}(b) presents the generalized squeezing parameter~\cite{toth07}
$\xi^2=\langle(\Delta\hat J^x)^2+(\Delta\hat J^y)^2+(\Delta\hat J^z)^2\rangle/(N/2)$
as a function of pump in the steady state for $N=25$. The TWA closely tracks the exact value as it decreases below unity, indicating entanglement~\cite{MA11} relevant for quantum information. The TWA also captures subradiance, absent in a mean-field description of PI ensembles. Here, destructive interference suppresses the normalized collective emission rate, $s=\langle\hat J^+\hat J^-\rangle/\langle\hat J^z+N/2\rangle$, below that of independent emitters ($s=1$), offering a distinct resource for metrology~\cite{facchinetti16,ostermann13}. For $\gamma=0.4\Gamma$, the upper inset confirms agreement throughout evolution and identifies subradiance as a dynamical precursor to squeezing~\cite{shankar21}. \\
\indent Dynamics beyond mean-field descriptions in PI ensembles are often approximated using a second-order cumulant expansion~\cite{plankensteiner22}, which neglects connected correlations beyond second order. In Fig.~\ref{fig:fig2}(b), the TWA more closely reproduces $\xi^2$ and normalized emission around their peaks and the upper lasing threshold $\gamma=N\Gamma$. Near this threshold, non-Gaussian correlations~\cite{barberena2025}, captured in the TWA, accompany an overshoot in cumulant predictions whose relative error increases with $N$, while the TWA approaches the exact result (lower inset). \\
\indent When the spin length is of order unity, the semiclassical expansion in Eq.~\eqref{eq:semiclassical-commutator} is no longer formally controlled, and TWA accuracy must be assessed empirically. In Fig.~\ref{fig:fig2}(b), the relative error in predicting squeezing changes from below $1\%$ at $\gamma=10\Gamma$ to approximately $10\%$ at $\gamma=0.01\Gamma$, as the system enters the deeply subradiant regime $\langle\hat J^+\hat J^-\rangle\lesssim 1$, reached for $\gamma\lesssim\Gamma/N$, where $\mathcal J=O(1)$ along individual trajectories. The non-Gaussian correlations~\cite{shankar21} present here again produce a larger relative error of approximately $29\%$ in the cumulant expansion, highlighting improved accuracy of the TWA even in the absence of formal semiclassical control. For single emitters with spin 1/2, accuracy has also been demonstrated for the TWA obtained in a path-integral formulation~\cite{hosseinabadi25}, whose equations of motion agree with those obtained using the present TWA for $N=1$ (Appendix).

\begin{figure}
    \centering
    \includegraphics[width=0.89\linewidth]{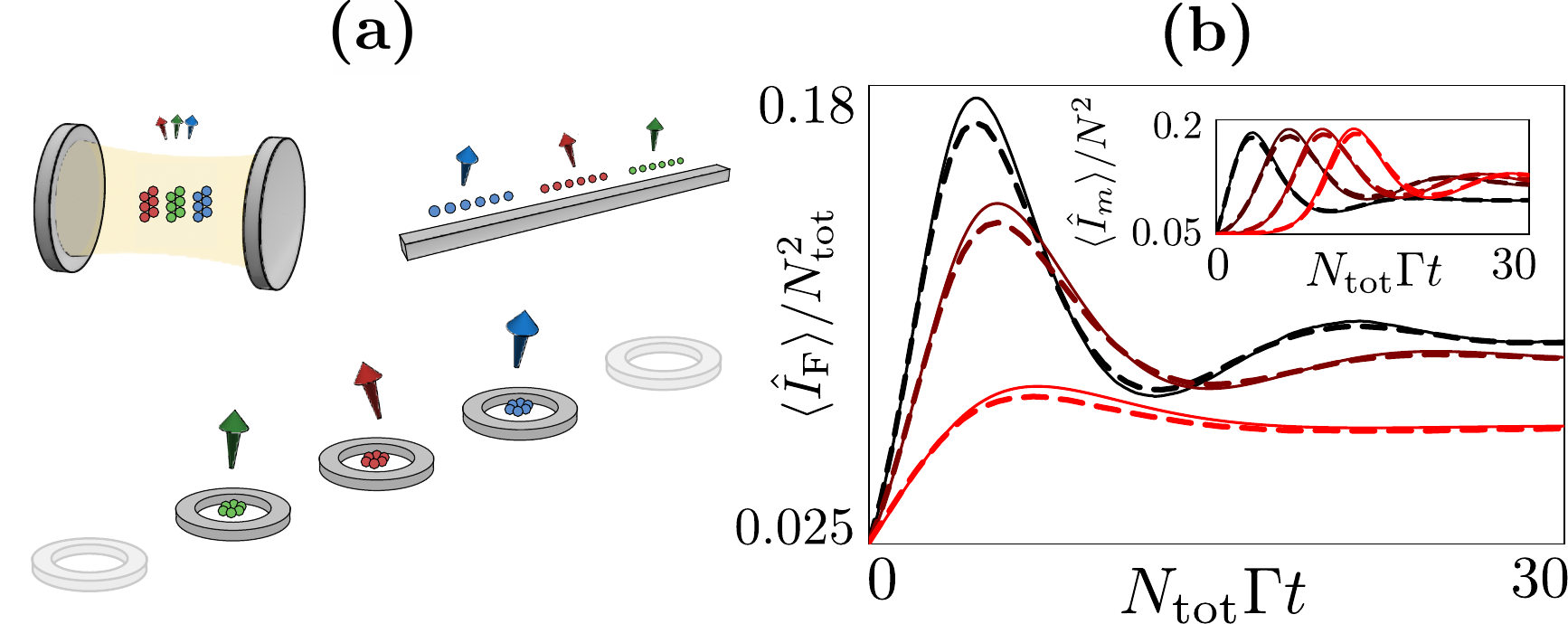}
    \caption{(a) Illustration of composite systems described using the TWA by assigning a distinct collective spin (arrow) to each PI emitter ensemble (dots). (b) Normalized forward emission rate $\langle \hat{I}_{\mathrm{F}}\rangle/N_{\mathrm{tot}}^2 = \langle \hat{L}_{\mathrm{F}}^{\dagger}\hat{L}_{\mathrm{F}}\rangle/N_{\mathrm{tot}}^2$ for $M=2$ pumped ensembles, each comprising $N=20$ emitters for a total $N_{\mathrm{tot}} = 40$ in the system Eq.~\eqref{eq:WQED}. The propagation phase $\varphi=0,\pi/4,\pi/2$ (upper to lower curves). Emitters are initially in the excited state, and the local pump $\gamma = 10\Gamma$. Inset: (identical) site-resolved emission rates $\langle \hat{I}_{m}\rangle/N^2 = \langle\hat J_m^+\hat J_m^-\rangle/N^2$ for $\varphi=0$ as the number of ensembles is increased with $M=1, 10, 10^2, 10^3$ (dark to light) and $\gamma = 0.25N_{\mathrm{tot}}\Gamma$. Solid (dashed) curves give the exact dynamics (TWA).}
    \label{fig:fig3}
\end{figure}

\begin{figure}
    \centering
    \includegraphics[width=\columnwidth]{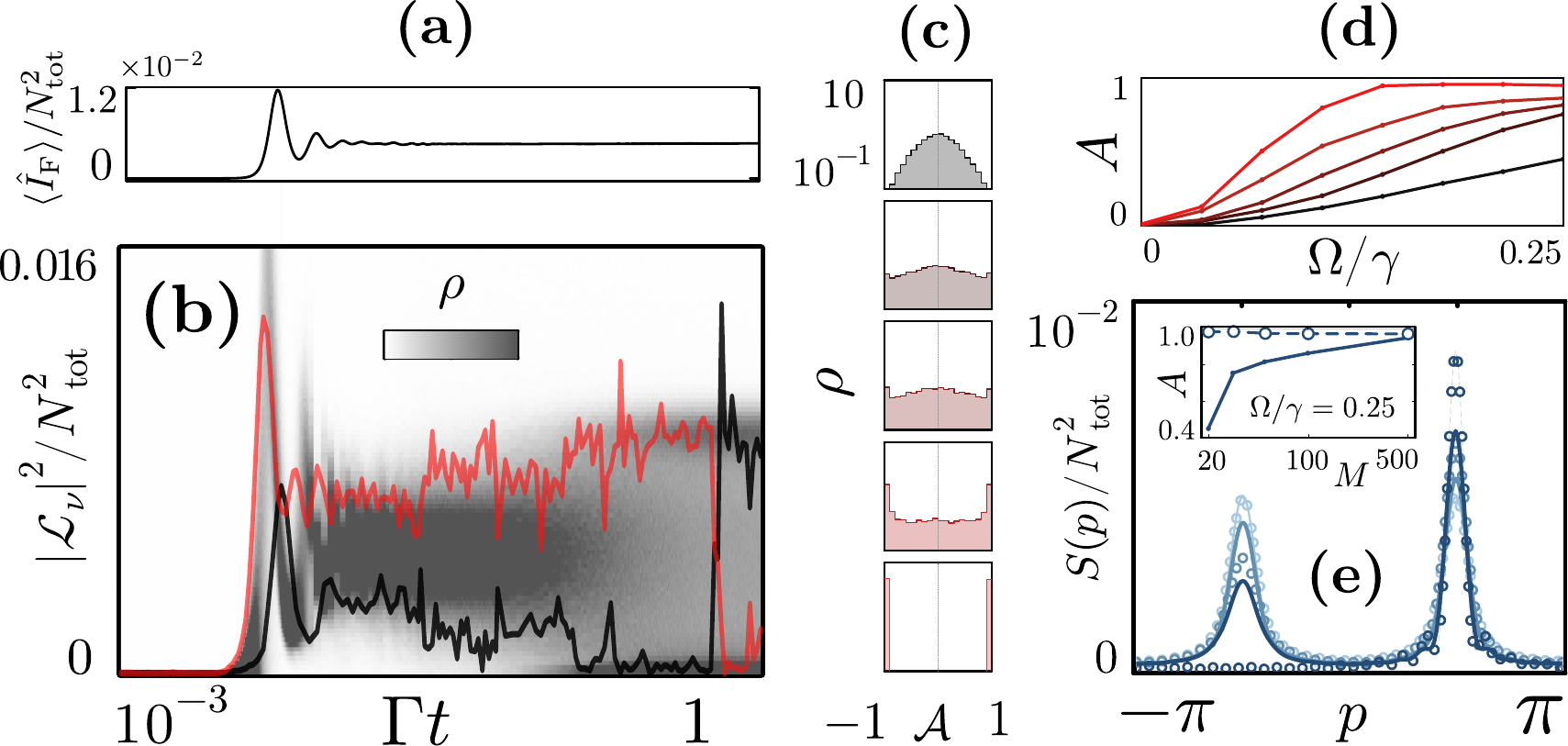}

    \caption{Emission dynamics in a 1D chain of ensembles [Eq.~\eqref{eq:WQED}], for emitters initially in the ground state. (a) Forward emission rate and (b) marginal distribution $\rho$ of the semiclassical signal power $|\mathcal{L}_{\nu}|^2$ in one output port, equal for $\nu = \mathrm{F},\mathrm{B}$, as a function of time. The forward and backward powers for an illustrative trajectory are denoted by black and red, respectively. (c) Distribution $\rho$ of trajectory-level directionality $\mathcal{A} = (|\mathcal{L}_{\mathrm{F}}|^2 - |\mathcal{L}_{\mathrm{B}}|^2)/(|\mathcal{L}_{\mathrm{F}}|^2 + |\mathcal{L}_{\mathrm{B}}|^2)$ over individual trajectories in the absence of a coherent drive, and (d) unconditional directionality $A = \langle \hat{I}_{\mathrm{F}} - \hat{I}_{\mathrm{B}}\rangle/\langle \hat{I}_{\mathrm{F}} + \hat{I}_{\mathrm{B}}\rangle$ at $t = 2/\Gamma$ versus normalized drive strength $\Omega/\gamma$ for $M=20,30,50,100,$ and $500$ (dark to light). In (c), $M$ increases from top to bottom. (e) Structure factor $S(p)$ at $t=2/\Gamma$ with $M=20$, for drives $\Omega/\gamma = 0,0.1,0.25$ (light to dark). Mean-field and TWA predictions are denoted by circles and solid lines, respectively. Inset shows predictions of $A$ at $\Omega/\gamma = 0.25$ for increasing $M$. In (c), (d), and (e), we fix the total propagation phase $M\varphi = 10\pi$ and effective cooperativity $\Gamma N_{\mathrm{tot}}/\gamma = 100$.}
    \label{fig:fig4}
\end{figure}

\indent In addition to a macroscopic spin length, formal control in semiclassical expansions requires the Wigner function to remain smooth over the effective Planck scale~\cite{zurek03}. For strongly nonlinear quantum dynamics in two-axis counter-twisting, we have confirmed~\cite{supplement} that rapid oscillations in the Wigner function at macroscopic $\mathcal J$ accompany deviations from exact dynamics for weak local decay, while agreement throughout dynamics is recovered once decay exceeds the twisting strength.\\
\indent These results establish a semiclassical limit for the quantum dynamics of PI emitter ensembles. The Wigner--Moyal expansion on the intrinsic classical phase space of a variable-length collective spin yields a valid stochastic semiclassical theory~\cite{supplement} that captures leading quantum corrections for arbitrary PI ensembles. It does not prescribe a closure of operator correlations and can describe diverse dynamical regimes beyond the steady state where low-order closures show more pronounced discrepancies~\cite{supplement,Zens2019,fasser25,stitely23}. \\
\indent \textbf{Spatially extended systems in the TWA}---Beyond individual homogeneous emitter ensembles, light--matter dynamics in spatially extended systems poses a central many-body problem in which microscopic fluctuations and spatial correlations shape macroscopic optical responses. By representing many such systems within this landscape---including waveguide-coupled emitter chains, arrays in cavities, and ensembles distributed across extended resonator lattices~[Fig.~\ref{fig:fig3}(a)]---as distinct, interacting PI ensembles, we apply the semiclassical expansion to obtain a phase-space description and a semiclassical limit for this broader class of systems (see Appendix). With an independent $\bm{\mathcal Z}_{m}$ assigned to each ensemble and evolved using Eq.~\eqref{eq:SDE}, dynamics can be described by a set of interacting variable-length spins, whose total number of phase-space variables scales linearly with the number of distinct ensembles. \\
\indent In \cite{supplement}, we have demonstrated an application of this TWA to the central spin model. To more broadly illustrate this approach here, we consider an infinite one-dimensional lattice of coupled resonators forming a waveguide. $M$ equally spaced sites are each uniformly coupled to $N$ emitters, for a total $N_{\mathrm{tot}} = NM$. We assume a sufficiently weak coupling so that the waveguide correlation time and photon transit time across the chain are both much shorter than the shortest timescale governing emitter evolution, and a Markovian description of the ensemble dynamics is valid. Eliminating the photonic degrees of freedom then yields an effective master equation for the emitter ensembles, with resonator-mediated coherent couplings and dissipative jumps: \begin{align}
    \hat{H} = \frac{\Gamma}{2} \! \sum_{\substack{m=1\\ m'=1}}^{M} \! \sin\left(\varphi|m{-}m'|\right)\hat{J}^{+}_{m}\hat{J}^{-}_{m'}, \ \hat{L}_{\mathrm{F}(\mathrm{B})}\!  = \! \sum_{m=1}^{M}e^{\pm i\varphi m}\hat{J}^{-}_{m}.
\label{eq:WQED}
\end{align}
Here, $\Gamma$ denotes the decay rate into the chain and $\varphi$ the propagation phase accumulated between neighboring ensembles. The jump operators $\hat{L}_{\mathrm{F}(\mathrm{B})}$  describe collective emission into forward- and backward-propagating modes, each at rate $\Gamma/2$. This setup realizes the waveguide-QED paradigm~\cite{sheremet23} with homogeneous ensembles acting as super-atoms. Applying the TWA yields a perturbative semiclassical theory for the spin model of 1D light propagation, recovering spatially discretized classical Maxwell–Bloch dynamics~\cite{lugiato2015} to leading order in $\mathcal J^{-1}$ in the deterministic thermodynamic limit. \\
\indent Coupling two ensembles already leads to qualitatively distinct dynamics~\cite{xu14,fasser24}, which may be described semiclassically. In Fig.~\ref{fig:fig3}(b), for $M=2,$ increasing the propagation phase suppresses collective emission under uniform local pumping. At $\varphi=\pi/2$, the TWA tracks a reduction of the emission peak by more than $50\%$ and of the steady-state emission by more than $30\%$, deviating markedly from the Dicke limit. At $\varphi=0$, comparing with Monte Carlo simulations, we capture site-resolved emission as $M$ increases up to one thousand [Fig.~\ref{fig:fig3}(b), inset], while separate comparisons confirm improving accuracy with increasing $N$. These results validate a semiclassical description of multiple ensembles. \\
\indent We now turn to the collective optical response of a large-scale medium in which strong dynamical quantum noise motivates the TWA. We consider a chain of $M=100$ ensembles, each containing $N=250$ emitters, for a total $N_{\mathrm{tot}}=25000$. We take finite $\varphi=\pi/10$, subject the emitters to uniform local pumping at $\gamma=250\Gamma$, and simulate the resulting dynamics using the TWA. In Fig.~\ref{fig:fig4}(a), a burst is observed in the forward emission rate $\langle \hat{I}_{\mathrm{F}}\rangle  = \langle\hat{L}_{\mathrm{F}}^{\dagger}\hat{L}_{\mathrm{F}}\rangle$, followed by steady-state superradiance. As expected, emission in the two directions is equal by mirror symmetry of the chain. \\
\indent Beyond unconditional expectations, the TWA offers a semiclassical description of individual output signals~\cite{leeRuostekoski14, vicentini18}. 
For the field exiting port $\nu = \mathrm{F, B}$, the semiclassical signal amplitude is given through input--output theory by the directional collective jump symbol, $\sqrt{\Gamma/2}\,\mathcal{L}_{\nu}$, appearing in the unraveling Eq.~\eqref{eq:SDE}. Adding the corresponding vacuum-noise realization along an individual trajectory then corresponds to a possible single-shot semiclassical heterodyne record~\cite{supplement}. \\
\indent In Fig.~\ref{fig:fig4}(b), an illustrative signal trajectory develops a directional imbalance under initial fluctuations that grows under uniform pumping until nearly all signal power occupies one direction~\cite{cardenas25}, with the distributions $|\mathcal{L}_{\nu}|^2$ showing low- and high-power branches. Fig.~\ref{fig:fig4}(c) confirms how directionality in collective emission emerges with increasing $M$, with the total propagation phase $M\varphi$ and effective cooperativity $\Gamma N_{\mathrm{tot}}/\gamma$ held constant. Crucially, macroscopic fluctuations in the signal due to dynamical quantum noise cause near-complete reversals of direction during evolution.  \\
\indent  To expose the effect of dynamical quantum noise in the unconditional optical response, we introduce a coherent waveguide drive, $\hat{H}_{\mathrm{drv}}=\Omega\sum_{m}(e^{i\varphi m}\hat{J}_{m}^{-}+\mathrm{h.c.})$. In Fig.~\ref{fig:fig4}(d), the unconditional directionality $A = \langle \hat{I}_{\mathrm{F}} - \hat{I}_{\mathrm{B}}\rangle/\langle \hat{I}_{\mathrm{F}} + \hat{I}_{\mathrm{B}}\rangle$ of total emission increases with $\Omega$ as the drive biases trajectories to the forward direction, for approximately constant total emission rate. This increase, sharpening with increasing $M$, is also captured by a mean-field theory of superradiance including initial fluctuations~\cite{haake79}. \\
In Fig.~\ref{fig:fig4}(e), however, the mean-field equations of motion produce qualitatively distinct predictions. In the unnormalized structure factor $S(p) = \sum_{m,m'}e^{-ip(m-m')}\langle \hat{J}^{+}_{m}\hat{J}^{-}_{m'}\rangle$, backward-propagating collective excitations, and hence backward emission $\langle \hat{I}_{\mathrm{B}}\rangle=S(-\varphi)$, fall almost to zero with increasing drive under classical evolution. The TWA retains an appreciable fraction while noise continues to drive reversals of emission direction in individual trajectories. These reversals lie outside a self-consistent linearized theory of small fluctuations about a single classical branch. As $M$ increases at $\Omega/\gamma = 0.25$, a large difference persists in predictions of the unconditional directionality in emission, with the classical prediction only close to the TWA at $M=500$, for more than one hundred thousand total emitters. \\
\indent \textbf{Conclusions}---We have established a semiclassical limit for the dynamics of a PI emitter ensemble with local dissipation. This construction, through a Wigner–Moyal expansion on the phase space of a variable-length spin, reconciles the Bloch-sphere phase-space description of collective spin dynamics with local dissipation. The resulting TWA is applicable across a broad class of quantum-optical models comprising PI emitter ensembles, improving upon the predictions of common methods based on prescribed closures of operator correlations. Our approach offers a starting point for incorporating higher-order effects of quantum interference~\cite{polkovnikov10}, and highlights symmetry-adapted phase spaces as a possible route toward semiclassical descriptions of a wider range of open spin systems~\cite{twyeffort22,davidson15,Zhu2019}.\\
\indent 
This TWA provides a unified and scalable semiclassical framework for modeling interacting emitter ensembles in mesoscopic architectures~\cite{lei23,periwal21,astner17,robinson24,kato19,run25,zou14}, mirroring the established TWA for bosonic systems. As highlighted here in cooperative emission, this description can stimulate further study into how strong fluctuations reshape active quantum systems~\cite{nadolny25} and can guide the exploration of quantum technologies~\cite{niedenzu18,montenegro23,robinson24,hymas25} powered by cooperative phenomena.

\begin{acknowledgments}
We acknowledge financial support, in part, from Moonshot R\&D, JST JPMJMS256E. We appreciate discussions with W. J. Munro, V. M. Bastidas, E. Aspropotamiti, A. S. S{\o}rensen, K. M{\o}lmer, J. Ruostekoski, and K. Azuma. 
\end{acknowledgments}

\input{main.bbl}

\section{Appendix}

In this Appendix, we summarize key conceptual aspects of the TWA for PI ensembles, while detailed derivations and supporting calculations are provided in \cite{supplement}.

\appendix
\noindent\textit{Extended phase space for PI ensembles}---In the TWA, we approximately sample the Wigner function of a PI ensemble using semiclassical trajectories evolving under Eq.~\eqref{eq:SDE} on the four-dimensional phase space $\bm{\mathcal Z}=(\phi,\theta,\psi,\mathcal J)$.
The symplectic structure on this phase space is specified by the canonical Poisson brackets
\begin{subequations}
\label{eq:poisson-brackets}
\begin{align}
    \{\phi,(\mathcal{J}+1/2)\cos\theta\} &= 1, \label{eq:poisson-collective} \\
    \{\psi,\mathcal{J}+1/2\} &= 1, \label{eq:poisson-local}
\end{align}
\end{subequations}
which correspond to the phase space of a linear rigid rotor~\cite{landau76} in a rotated frame~\cite{Romero2015}.
As in the conventional description of a collective spin, the angles $\phi$ and $\theta$ specify the orientation of an angular-momentum-like vector of length $\mathcal J+1/2$.
In the absence of local dissipation, $\mathcal J$ is conserved and Eq.~\eqref{eq:poisson-collective} alone describes the reduced dynamics on the Bloch sphere.
With local dissipation, $\mathcal{J}$ becomes dynamical, and Eq.~\eqref{eq:poisson-local} identifies $\psi$ as its canonically conjugate angle~\cite{barberena2025,tomatani2015}. The local jumps $\ell^{jq}$ in Eq.~\eqref{eq:SDE} then couple $\psi$ and $\mathcal J$ to each other and to the spherical angles $(\phi,\theta)$. The dynamics closes on this four-dimensional phase space, which constitutes a minimal symplectic extension of the Bloch sphere sufficient to semiclassically describe local dissipation in PI ensembles. The resulting equations of motion can be integrated directly in these intrinsic variables, but for numerical convenience we use an equivalent extrinsic Cartesian parametrization~\cite{supplement}.

\noindent\textit{TWA for composite systems}---In Figs.~\ref{fig:fig3} and \ref{fig:fig4}, we applied the TWA to composite systems of $M$ distinct ensembles, whose interactions preserve permutation invariance within each ensemble. To extend the quantum--classical mapping to this case, we adopt the angular-momentum representation for each ensemble and take the full Stratonovich--Weyl kernel as a tensor product of single-ensemble kernels~\cite{tilma2016},
$\hat{\Delta}_{\mathrm{tot}}=\bigotimes_{m=1}^M \hat{\Delta}(\bm{\mathcal{Z}}_{m})$,
where $\hat{\Delta}(\bm{\mathcal Z}_m)$ acts on ensemble $m$. The semiclassical expansion then proceeds as in the single-ensemble case, with trajectories of ensemble $m$ evolving under Eq.~\eqref{eq:SDE} for variables $\bm{\mathcal Z}_m=(\phi_m,\theta_m,\psi_m,\mathcal J_m)$ and with Poisson brackets vanishing between variables of distinct ensembles. Local dissipation acts independently and identically on the emitters within each ensemble, so that jumps $\hat l^{jq}_m$ for ensemble $m$ contribute Weyl symbols $\mathcal{L}_k=\ell^{jq}(\bm{\mathcal Z}_m)$, with independent complex white noise for each channel $(m,j,q)$. Since operators on different ensembles commute, the Weyl symbol of the Hamiltonian in Eq.~\eqref{eq:WQED} is obtained by simply substituting each collective operator that appears with its Weyl symbol. The same substitution applies to collective jump operators, which share a single complex white-noise process per collective channel. The number of phase-space variables and noise processes then increases linearly with the number of distinct ensembles, and trajectories in Eq.~\eqref{eq:SDE} can be solved in parallel. With the evolution of semiclassical trajectories specified, we calculate composite observables such as $\langle \hat L_{\mathrm F}^\dagger \hat L_{\mathrm F}\rangle$ and $S(p)$, by decomposing into inter-ensemble products ($m\neq m'$) and on-site terms ($m=m'$). We use products of Weyl symbols to calculate the former and the exact Weyl symbol of $\hat J_m^{+}\hat J_m^{-}$~\cite{supplement} for the latter. \\
\indent Many composite systems can be partitioned directly into distinct PI ensembles, allowing the semiclassical expansion to be applied to each ensemble. More generally, inhomogeneous systems can be modeled by partitioning emitters into approximately homogeneous clusters $m$ of size $N_m$, as in approaches based on the cumulant expansion~\cite{lei23,Zens2019}. Within a cluster treated as PI, differences in transition frequencies, couplings, and local rates should remain dynamically unresolved for the observables and timescale of interest. The extent of the corresponding homogeneous region and the density of emitters within it then determine $N_m$. The partition is refined, typically reducing $N_m$, until the observables of interest converge.
Although the semiclassical expansion is not formally controlled at small $N_m$, the TWA can still capture dynamics in smaller clusters, with accuracy assessed empirically. This is observed in the example of Fig.~\ref{fig:fig2}(a) at $N = 3$, and we have separately confirmed similar qualitative agreement for $N_m = N =3$ in the model of Fig.~\ref{fig:fig3}. For clusters with $N_m=1$, local dissipation operators are identified with their collective counterparts, and the single-emitter SW kernel $\hat{\Delta}(\phi_m,\theta_m)$ is constrained to the $J=1/2$ spin sector. In \cite{supplement}, we show how dynamics including these clusters can still be described using the semiclassical Langevin equations in Eq.~\eqref{eq:SDE}, and demonstrate an application to the central spin model. In particular, we can recover the TWA benchmarked in Ref.~\cite{hosseinabadi25} when $N_m = 1$ for all $m$, unifying the TWA for larger PI clusters and individually resolved emitters.

\let\addtocontents\arxivAddToContents
\clearpage

\appendix
\onecolumngrid

\setcounter{equation}{0}
\setcounter{table}{0}
\renewcommand{\theequation}{S\arabic{equation}}
\renewcommand{\thetable}{S\arabic{table}}
\renewcommand{\theHequation}{supp.\arabic{equation}}
\renewcommand{\theHfigure}{supp.\arabic{figure}}
\renewcommand{\theHtable}{supp.\arabic{table}}
\renewcommand{\theHsection}{supp.\arabic{section}}
\renewcommand{\theHsubsection}{supp.\arabic{section}.\arabic{subsection}}
\renewcommand{\theHsubsubsection}{supp.\arabic{section}.\arabic{subsection}.\arabic{subsubsection}}

\setcounter{figure}{0}
\renewcommand{\thefigure}{\textbf{S\arabic{figure}}}

\renewcommand{\citenumfont}[1]{S#1}

\renewcommand{\bibnumfmt}[1]{[S#1]}

\begin{center}
{\large\bfseries Supplemental Material: Semiclassical Phase-Space Dynamics of Emitter Ensembles with Local Dissipation\par}
\vspace{0.5em}
L. Ruks\par
\vspace{0.25em}
{\small Basic Research Laboratories \& NTT Research Center for Theoretical Quantum Information, NTT, Inc., 3-1 Morinosato Wakamiya, Atsugi, Kanagawa, 243-0198, Japan\par}
\end{center}

In this Supplemental Material, we provide extended derivations and further benchmarks for this TWA, and provide additional details of its relation to existing methods.

\tableofcontents

\subsection{Stratonovich--Weyl kernel in the angular-momentum representation and calculation of the Weyl symbols}
In the Main Text, we obtain the truncated Wigner approximation (TWA) for Eq.~(1) using a Stratonovich--Weyl (SW) kernel in the angular-momentum representation for permutation-invariant (PI) emitter ensembles. Here, we briefly recap the angular-momentum representation before presenting the SW kernel and calculation of the Weyl symbols. 
\subsubsection{Angular-momentum representation}
The master equation in Eq.~(1) of the Main Text describes the evolution of $N$ two-level emitters, where states generally span a Hilbert space structured as an $N$-fold tensor product of each of the emitter Hilbert spaces. However, permutation invariance restricts the full evolution to a reduced $\mathrm{SU}(2)$-covariant subspace, which allows us to employ an efficient angular-momentum representation of dynamics. In this representation, we can then use a symmetry-adapted SW kernel~\cite{supp-klimov2008} to obtain a minimal quantum--classical mapping for PI systems. For the dynamics described by Eq.~(1), any PI density matrix can be expressed~\cite{supp-shammah2018,supp-chase2008},
\begin{equation}
\hat{\rho} = \sum_{JMM'}\rho_{JMM'}\times \frac{1}{d_{J}}\sum_{\alpha}|JM\alpha\rangle \langle JM'\alpha| = \sum_{JMM'}\rho_{JMM'}\overline{|JM \rangle \langle JM'|}.
\label{supp-eq:density}
\end{equation}
Here, the Dicke states $|JM\alpha\rangle$ are eigenstates of total spin squared $\hat J^{2}$ and inversion $\hat J^{z}$ for eigenvalues $J(J+1)$ (with nonnegative $J = N/2,\, N/2-1,\, \ldots$) and $M = -J,\ldots J$ respectively, with $d_{J}$ degeneracies indexed by $\alpha$. In particular, $\hat{\rho}$ is proportional to the identity on the multiplicity space, and so the coefficients $\rho_{JMM'}$ completely specify the density matrix through populations and inversion coherences within each total-spin sector $J$. Comparing with Eq.~(2) of the Main Text, a one-to-one correspondence can be identified between degeneracy-averaged Dicke-state outer products in the original space and outer products in the angular-momentum representation:
\begin{equation}
\overline{|JM \rangle \langle JM'|} \longleftrightarrow |JM\rangle\langle JM'|,
\label{supp-eq:collective-basis}
\end{equation}
for the collective kets $|JM\rangle$ defined in the Main Text. The action of collective operators $\hat{J}^{q}$ on elements $|JM\rangle\langle JM'|$ is then consistent with that on $\overline{|JM\rangle\langle JM'|}$ in the original Dicke-state basis. To demonstrate that the aggregate action of local dissipation in the angular-momentum representation is also consistent with the underlying dynamics, we note that the effect of PI local dissipation in Eq.~(1) of the Main Text, for each $q$, is to map $\overline{|JM\rangle\langle JM'|}$ to a linear combination~\cite{supp-chase2008,supp-kolmer2024,supp-shammah2018} of $\overline{|J M\rangle\langle J M'|}$ and $\overline{|J+j \  M+q\rangle\langle J+j \ M'+q|}$, summed over $j = -1,0,1$. In the angular-momentum representation, the corresponding $|JM\rangle\langle JM'|$ are mapped to the same linear combination of $|J  M\rangle\langle J M'|$ and $|J+j \ M+q\rangle\langle J+j \ M'+q|$, i.e., Eq.~(3) of the Main Text is satisfied, for the jump operators~\cite{supp-kolmer2024},
\begin{equation}
\hat {l}^{jq} \, := \sum_{J'}\Lambda^{j,q}_{J',N}\hat{T}_{1,q}^{J'+j,J'}.
\label{supp-eq:collective-L}
\end{equation}
In the angular-momentum representation, the generalized spherical tensors are defined by
\begin{equation}
\hat{T}_{k,q}^{J',J} = \sqrt{\frac{2k+1}{2J'+1}}
\sum_{M,M'} C^{J',M'}_{k,q;J,M}|J'M'\rangle\!\langle JM|,
\label{supp-eq:gen-tensors}
\end{equation}
for Clebsch--Gordan coefficients $C^{\,J',M'}_{\,k,q;\,J,M}$.
For dephasing ($q=0$), pumping ($q=1$), and decay ($q=-1$), the coefficients are
\begin{equation}\label{supp-eq:Lambda}
\Lambda^{\,j,q}_{J',N}=c_{j}c_{q}\sqrt{\frac{\left[N+2-j(2J'+j+1)\right]\,\bigl(2J'+2j+1\bigr)}{6}},
\end{equation}
with
\begin{equation}
c_j=\begin{cases}-1,& j=1\\ +1,& j=0,-1\end{cases},\qquad
c_q=\begin{cases}q,& q=\pm1\\ -\sqrt2,& q=0\end{cases}
\end{equation}
We note that the effective jumps $\hat{l}^{jq}$ do not represent physically accessible decay channels, but are mathematical constructions that reproduce the correct system dynamics in the symmetry-adapted representation. This derivation shows that the dynamics of Eq.~(1) are exactly reproduced in the angular-momentum representation, for the jumps $\hat{l}^{jq}.$

\subsubsection{Stratonovich--Weyl kernel and calculation of Weyl symbols}

With the angular-momentum representation identified, we can apply an SW kernel developed for variable-spin systems~\cite{supp-klimov2008,supp-klimov17}. The SW kernel $\hat\Delta(\bm{\mathcal Z})=\hat U(\bm{\Omega})\hat\Pi_{\mathcal J}\hat U^\dagger(\bm{\Omega})$ is built from collective rotations $\hat{U}(\bm{\Omega}) = e^{-i\hat{J}_{z}\phi}e^{-i\hat{J}_{y}\theta}e^{-i\hat{J}_{z}\psi}$, for the Euler angles $\bm{\Omega} = (\phi,\theta,\psi)$, acting on the parity operator $\hat{\Pi}_{\mathcal{J}}$,
\begin{equation}\label{supp-eq:gen-SW}
\hat\Pi_{\mathcal J} = \sum_{{K=0}}^{2\mathcal{J}}\;\sum_{Q=-K}^{K}
\sqrt{\frac{2K+1}{2\mathcal{J}+1}}
\hat T_{K,Q}^{\mathcal{J}+\frac{Q}{2},\mathcal{J} - \frac{Q}{2}},
\end{equation}
where the spherical tensors contain products $|JM\rangle \langle J'M'|$ for $J \neq J'$ and can capture changes in total spin caused by local dissipation. Here, we write $\mathcal{J} = \mathrm{j}/2$ for the integer $\mathrm{j}$ used in \cite{supp-klimov2008,supp-klimov17}, to identify $\mathcal{J}$ as a classical spin-like index. This SW kernel establishes a one-to-one correspondence between operators $\hat{O}$ and Weyl symbols $\mathcal{O} = \mathrm{Tr}[\hat{O}\hat{\Delta}]$, satisfying the usual axioms required for the Wigner function $W$ to define a quasiprobability distribution~\cite{supp-klimov2008}. To calculate Weyl symbols, it suffices to express operators as sums of generalized spherical tensors. Using the relation
\begin{equation}
\hat U(\bm{\Omega})\,\hat T_{K,Q}^{J',J}\,\hat U^\dagger(\bm{\Omega})
= \sum_{Q'=-K}^{K} D^{K}_{Q'Q}(\bm{\Omega})\,\hat T_{K,Q'}^{J',J},
\end{equation}
we express the kernel explicitly in the tensor basis. For operators $\hat{O}$ also expressed in this basis, taking the trace $\mathrm{Tr}[\hat{O}\hat{\Delta}]$ and exploiting orthonormality, $\mathrm{Tr}[(\hat{T}^{J',J}_{k,q})^{\dagger}\hat{T}^{J'',J'''}_{\kappa,q'}]
=\delta_{J,J'''}\delta_{J',J''}\delta_{k,\kappa}\delta_{q,q'}$, then gives the Weyl symbol $\mathcal{O}$. For the local jump operators $\hat{l}^{jq}$ introduced in the Main Text, calculating the Weyl symbols $\ell^{jq} = \mathrm{Tr}[\hat{l}^{jq}\hat{\Delta}]$ then yields Eq.~(4) for the jump amplitudes,
\begin{equation}\label{supp-eq:chi-corrected}
\chi_{jq}(\mathcal{J})=c_{j}c_{q}\sqrt{\frac{\bigl[N+2-j(2\mathcal{J}+1)\bigr]\bigl(2\mathcal{J}+j+1\bigr)}{2(2\mathcal{J}+1)}}.
\end{equation}
Note that these amplitudes vanish for $j=-1, \mathcal{J} = 0$, and $j=1,\mathcal{J}=(N+1)/2.$ For the collective operators, a similar calculation gives
\begin{equation}
\chi_q(\mathcal{J})=\tilde{c}_{q}\sqrt{2\mathcal{J}\left(\mathcal{J}+1\right)}, \qquad \tilde{c}_q=\begin{cases}-q,& q=\pm1\\ 1/\sqrt2,& q=0\end{cases}
\end{equation}
yielding the Weyl symbols of one-body operators $\hat{J}^{q}$ used in the Main Text. To obtain the Weyl symbols used in two-body expectations $\langle \hat{J}^{+}\hat{J}^{-}\rangle$, we note the decomposition $\hat{J}^{+}\hat{J}^{-} = (\hat{J}^{x})^2 + (\hat{J}^{y})^2 + \hat{J}^{z}$ into a sum of symmetrized products, and use the Weyl symbol~\cite{supp-klimov2008} for symmetrized quadratic operators ($\mu,\nu = x,y,z$):
\begin{equation}
      \quad \mathrm{Tr}\left[\left(\frac{\hat{J}^{\mu}\hat{J}^{\nu} + \hat{J}^{\nu}\hat{J}^{\mu}}{2}\right)\hat{\Delta}\right] = \left[\frac{\mathcal{J}^{\mu}\mathcal{J}^{\nu}}{\mathcal{J}(\mathcal{J}+1)} - \delta_{\mu \nu}/3\right]\sqrt{\mathcal{J}\left(\mathcal{J}-\frac{1}{2}\right)\left(\mathcal{J}+\frac{3}{2}\right)\left(\mathcal{J}+1\right)} + \mathcal{J}(\mathcal{J}+1)\delta_{\mu\nu}/3,
\end{equation}
where $\mathcal{J}^{\nu}$ is the Weyl symbol of $\hat{J}^{\nu}.$ These expressions provide the Weyl symbols used in the stochastic differential equations and evaluation of observable expectations in the Main Text.

\subsection{Derivation of the semiclassical stochastic differential equations}
In the Main Text, we used the stochastic differential equation Eq.~(6) to simulate the dynamics of Eq.~(1) in the truncated Wigner approximation. Here, we outline how Eq.~(6) is obtained using the SW kernel, in the angular-momentum representation, following the derivations in \cite{supp-dubois2021,supp-Walter98}. We trace Eq.~(1) against the SW kernel to obtain the evolution of the Wigner function in a Moyal equation. The Moyal equation cannot be solved in general, but using the semiclassical expansion, Eq.~(5), developed in \cite{supp-klimov2008,supp-tomatani2015}, we approximately map operator commutators to Poisson brackets. Neglected terms are organized in powers of $\mathcal{J}^{-1}$ and include higher-order derivatives acting on the Weyl symbols. Equivalently, the expansion can be organized in powers of $\hbar$, treated as a formal parameter controlling the strength of quantum fluctuations~\cite{supp-dubois2021,supp-polkovnikov10}. Applying Eq.~(5) in the Moyal equation retains the leading quantum corrections~\cite{supp-hosseinabadi25}, giving the second-order partial differential equation for the Wigner function $W$,
\begin{align}
\partial_t W &= \{\mathcal{H},W\}
+\frac{i}{2}\sum_k r_{k}\Big(\{\mathcal{L}_k W,\mathcal{L}_k^\ast\}+\{\mathcal{L}_k, W \mathcal{L}_k^\ast\}\Big) -\frac{1}{4}\sum_k r_{k}\Big(\{\{\mathcal{L}_k,W\},\mathcal{L}_k^\ast\}+\{\mathcal{L}_k,\{W,\mathcal{L}_k^\ast\}\}\Big),
\label{supp-eq:wigner-evolution}
\end{align}
with $\mathcal{L}_{k}, r_{k}$ as in the Main Text, and $k$ indexing jump channels. Here, the Poisson bracket in canonical coordinates is:
\begin{equation}
    \{f,g\}
= \frac{\partial f}{\partial \phi}\frac{\partial g}{\partial p_{\phi}}
- \frac{\partial f}{\partial p_{\phi}}\frac{\partial g}{\partial \phi}
+ \frac{\partial f}{\partial \psi}\frac{\partial g}{\partial p_{\psi}}
- \frac{\partial f}{\partial p_{\psi}}\frac{\partial g}{\partial \psi},
\end{equation}
for
\begin{equation}
    p_{\phi} = \cos\theta (\mathcal{J}+1/2), \quad p_{\psi} = (\mathcal{J}+1/2),
\end{equation}
as outlined in the Appendix.  Eq.~\eqref{supp-eq:wigner-evolution} is covariant under coordinate transformation and, in these canonical coordinates $\bm{\mathcal{Z}}'$, can be mapped to a Fokker--Planck equation:
\begin{equation}
\label{supp-eq:FPE}
\frac{\partial}{\partial t}W(\bm{\mathcal{Z}}',t)
+\sum_{i}\frac{\partial}{\partial \mathcal{Z}_i'}\!\big[u_i(\bm{\mathcal{Z}}')\,W(\bm{\mathcal{Z}}',t)\big]
-\frac{1}{2}\sum_{ij}\,\frac{\partial^2}{\partial \mathcal{Z}_i'\,\partial \mathcal{Z}_j'}\!\big[D_{ij}(\bm{\mathcal{Z}}')\,W(\bm{\mathcal{Z}}',t)\big]
=0.
\end{equation}
The diffusion matrix reads
\begin{equation}\label{supp-eq:diff-matrix}
D_{ij}(\bm{\mathcal{Z}}',t)
= \frac{1}{2}\sum_k r_{k}\big( \{\mathcal{Z}_i',\mathcal{L}_k\}\{\mathcal{Z}_j',\mathcal{L}_k^\ast\} + \{\mathcal{Z}_i',\mathcal{L}_k^\ast\}\{\mathcal{Z}_j',\mathcal{L}_k\} \big),
\end{equation}
and the drift is
\begin{align}   u_i(\bm{\mathcal{Z}}',t) = \{\mathcal{Z}_i',\mathcal{H}\}
- \frac{i}{2} \sum_k r_{k}\big( \mathcal{L}_k\,\{\mathcal{Z}_i',\mathcal{L}_k^\ast\} + \mathcal{L}_k^\ast\,\{\mathcal{L}_k,\mathcal{Z}_i'\} \big)
- \frac{1}{4}\sum_k r_{k}\big( \{\{\mathcal{L}_k,\mathcal{Z}_i'\},\mathcal{L}_k^\ast\} + \{\{\mathcal{L}_k^\ast,\mathcal{Z}_i'\},\mathcal{L}_k\} \big).
\end{align}

This Fokker--Planck equation can be unraveled into an It\^o SDE and then converted to Stratonovich form, yielding Eq.~(6) of the Main Text. As the Stratonovich SDE is also covariant under coordinate transformation, it may be interpreted in the presented phase-space coordinates $\bm{\mathcal{Z}} = (\phi,\theta,\psi,\mathcal{J})$.

We here note that quantum noise correlations for nonlinear dissipation need not admit a direct interpretation in terms of classical diffusion. Consequently, Fokker--Planck phase-space equations obtained for spin ensembles can have negative diffusion and can therefore fail to admit a real stochastic unraveling~\cite{supp-olsen05,supp-carmichael2008,supp-merkel20,supp-huber22}. Discarding diffusion terms to restore positivity can remove relevant contributions~\cite{supp-huber21}, affecting accuracy particularly at small spin length or near critical dynamics [see Fig.~\ref{supp-fig:schwinger}]. In the present TWA, the diffusion in Eq.~\eqref{supp-eq:diff-matrix},  derived using the semiclassical Wigner--Moyal expansion, takes exactly the form of classical Hamiltonian diffusion on phase space, with stochastic increments generated through Poisson brackets of the Weyl jump symbols. Their covariance gives a diffusion matrix that is positive semidefinite by construction, yielding a valid stochastic description for arbitrary PI ensembles that consistently retains the leading quantum corrections.

\subsubsection{Recovering collective dynamics}

When the dynamics are purely collective, all Weyl symbols in the Lindbladian are independent of $\psi$, so $\mathcal{J}$ is constant. The dynamics then reduce~\cite{supp-klimov17} to the two-dimensional Bloch sphere $(\phi,\theta)$. Neglecting relative $O(\mathcal{J}^{-2})$ corrections in the Poisson bracket,
\begin{equation}
    \{\mathcal{J}^{\alpha},\mathcal{J}^{\beta}\} = \left(1 + O(\mathcal{J}^{-2})\right)\epsilon_{\alpha \beta \gamma}\mathcal{J}^{\gamma},
\end{equation}
recovers the standard spin Poisson algebra and the Fokker--Planck equation established in a Wigner--Moyal semiclassical expansion for collective spin dynamics~\cite{supp-dubois2021}. When the system is closed, $\mathcal{L}_{k} = 0$ for all $k$, trajectories are deterministic and follow, up to relative order $\mathcal{J}^{-2}$ corrections, the mean-field equations of motion~\cite{supp-altland09} for Hamiltonian dynamics. \\

\subsubsection{Recovering the TWA for individual emitters}

\indent  While the present TWA is derived in the formal semiclassical limit, the TWA for single emitters, derived in a path-integral formulation~\cite{supp-hosseinabadi25}, can be heuristically recovered in the case of individual emitters as mentioned in the Appendix. To this end, we identify local dissipative operators with their corresponding collective operators. Applying the semiclassical expansion and taking the standard spin Poisson algebra, $\{\mathcal J^\alpha,\mathcal J^\beta\}=\epsilon_{\alpha\beta\gamma}\mathcal J^\gamma$, yields Eq.~\eqref{supp-eq:FPE} for fixed spin $J=\mathcal{J} =1/2$ and the corresponding semiclassical Langevin equations, Eq.~(6) of the Main Text, on the Bloch sphere. For a single emitter, these are equivalent to those of Ref.~\cite{supp-hosseinabadi25}, with Weyl symbols $\mathcal{J}^{\alpha}$ satisfying $(\mathcal{J}^{x})^2+(\mathcal{J}^{y})^2+(\mathcal{J}^{z})^2=3/4$. The extension to multiple PI clusters then follows as noted in the Appendix. We finally note that this TWA can be formally derived through a more careful consideration of the star product in the Moyal equation. In particular, retaining all SU$(2)$ corrections
yields the exact evolution for a single emitter~\cite{supp-noel2026}.

\subsection{Evolution in extrinsic Cartesian coordinates}

In the Main Text, we formulated the semiclassical dynamics in the intrinsic coordinates
$\bm{\mathcal Z}=(\phi,\theta,\psi,\mathcal J)$ of a linear rigid rotor.
For numerical integration it is convenient to work in an equivalent set of extrinsic Cartesian variables,
which avoids coordinate singularities and allows a simple implementation. We introduce an angular-momentum-like vector~\cite{supp-tomatani2015}
\begin{equation}
\bm{\mathcal J}:=(\mathcal{J}+1/2)\,\hat{\mathbf{n}}(\theta,\phi),
\end{equation}
where $\hat{\mathbf n}(\theta,\phi)$ is the usual unit vector on the sphere. We note that the components $\mathcal{J}_{\alpha}$ of $\bm{\mathcal{J}}$ agree with the Weyl symbols $\mathcal{J}^{\alpha}$ up to relative corrections $O(\mathcal{J}^{-2})$. We obtain the orthonormal basis
\begin{align}
\hat{\mathbf{n}}(\theta,\phi) &= (\sin\theta\cos\phi,\ \sin\theta\sin\phi,\ \cos\theta),\\
\hat{\mathbf{e}}_\theta(\theta,\phi) &= (\cos\theta\cos\phi,\ \cos\theta\sin\phi,\ -\sin\theta),\\
\hat{\mathbf{e}}_\phi(\theta,\phi) &= (-\sin\phi,\ \cos\phi,\ 0).
\end{align}
The auxiliary angle $\psi$ is encoded by a unit vector in the tangent plane,
\begin{equation}
\bm{\mathcal{R}}
= -\sin\psi\,\hat{\mathbf{e}}_\theta(\theta,\phi)
+ \cos\psi\,\hat{\mathbf{e}}_\phi(\theta,\phi),
\end{equation}
giving the constraints
\begin{equation}
\bm{\mathcal{R}}^2=1,\qquad \bm{\mathcal{J}}\cdot\bm{\mathcal{R}}=0.
\end{equation}
Using the intrinsic Poisson brackets in Eqs.~(9) of the Main Text, we verify the expected relations
\begin{equation}
\{\mathcal{R}_i,\mathcal{R}_j\}=0,\qquad
\{\mathcal{J}_i,\mathcal{R}_j\}=\epsilon_{ijk}\mathcal{R}_k,\qquad
\{\mathcal{J}_i,\mathcal{J}_j\}=\epsilon_{ijk}\mathcal{J}_k,
\end{equation}
Because the Stratonovich SDEs of Eq.~(6) of the Main Text are covariant under changes of variables, we equivalently numerically integrate the dynamics directly in the
$(\bm{\mathcal J},\bm{\mathcal R})$ coordinates.
In practice, we address deviations from the constraints due to a finite numerical time step by renormalizing $\bm{\mathcal R}$
and projecting out any accumulated component parallel to $\bm{\mathcal J}$ after each step. In addition, we apply a reflecting boundary condition at $\mathcal{J}=0$ and $\mathcal{J}=(N+1)/2$, where $\chi_{-1q}(\mathcal{J})$ and $\chi_{1q}(\mathcal{J})$ vanish respectively. The numerical integration is performed using \cite{supp-rackauckas17}. As observed in Fig.~2(b) of the Main Text, the simulations are generally capable of covering a broad range of dynamics, but can be computationally prohibitive, compared to cumulant expansions, for very large or small dissipation.

\subsection{Sampling the initial state Wigner function}

In the Main Text, we considered evolution of Eq.~(1) beginning from either the ground state or the completely inverted state, $|N/2 \ \pm \!\! N/2\rangle.$ Here, we outline how we sample initial conditions for semiclassical trajectories in the TWA. We present the result for a single ensemble, noting that for composite systems the phase-space variables for each distinct ensemble are sampled from the corresponding single-ensemble distribution. The Wigner function for the fully inverted state is
\begin{equation}
W_N(\bm{\mathcal{Z}}) = 
\frac{\delta_{\mathcal{J},N/2}}{8\pi^2}\sum_{K=0}^{N}(K+1/2)\,
(-1)^{K} C_{K,0;N/2,N/2}^{N/2,N/2}\,P_K(\cos\theta),
\end{equation}
where $P_{K}$ is the Legendre polynomial of order $K$. The Kronecker delta $\delta_{\mathcal{J},N/2}$ restricts the initial ensemble to the maximal-spin manifold $\mathcal{J}=N/2$. Because the initial state has a definite total spin, the conjugate angle $\psi$ is uniformly distributed and sampled over $[0,4\pi)$, while the azimuthal angle $\phi$ is sampled uniformly on $[0,2\pi)$. For $N\gtrsim 10$, quantum fluctuations of the fully inverted state are well captured by a Gaussian distribution~\cite{supp-polkovnikov10} in the polar angle,
\begin{equation}
W_{N}(\bm{\mathcal{Z}})=\frac{1}{\mathcal{N}}\exp\!\left(-N\theta^{2}/2\right)\delta_{\mathcal{J},N/2},
\label{supp-eq:initial-wigner}
\end{equation}
with $\mathcal{N}$ enforcing normalization. The initial distribution for the ground state follows upon setting $\theta \mapsto \pi-\theta$. For the $N=3$ benchmark in Fig. 2(a) of the Main Text and the convergence benchmarks with increasing emitter number (inset), we use the exact Wigner function to obtain a more representative distribution of the initial state. Here, we restrict $W_N(\theta)$ to $\theta\in[0,L]$, where $L$ is the smallest positive root of $W_N$, and renormalize to obtain a positive probability distribution from which we sample. Any remaining negative contributions are negligible already for moderate $N \gtrsim 10$~\cite{supp-davis21}, and this procedure yields the benchmark agreement shown in the Main Text. For the cases generally considered in Figs.~2--4, dynamics under the Gaussian approximation Eq.~\eqref{supp-eq:initial-wigner} closely match those obtained using the exact Wigner function. We note that using a gauge freedom in the Wigner function for small $N$ (see below) could enable an exact sampling prescription.

For individual emitters, $N=1$, the Wigner function of the initial state can be sampled exactly. Although the continuous Wigner function of a fully polarized spin-$1/2$ state contains negative regions, adding spherical harmonics with degree greater than one does not change the initial density matrix due to a gauge freedom for $J=1/2$~\cite{supp-mink22}. A discrete sampling for the Cartesian coordinates $\mathcal J^x,\mathcal J^y=\pm1/2$ independently with equal probabilities, with fixed $\mathcal J^z=\pm1/2$, equivalently represents the initial density matrix and provides valid initial conditions for semiclassical evolution, used in Fig.~\ref{supp-fig:central-spin}.

\subsection{Local dephasing and decay in a boundary time crystal}
\begin{figure}[h!]
    \centering
    \includegraphics[width=0.65\linewidth]{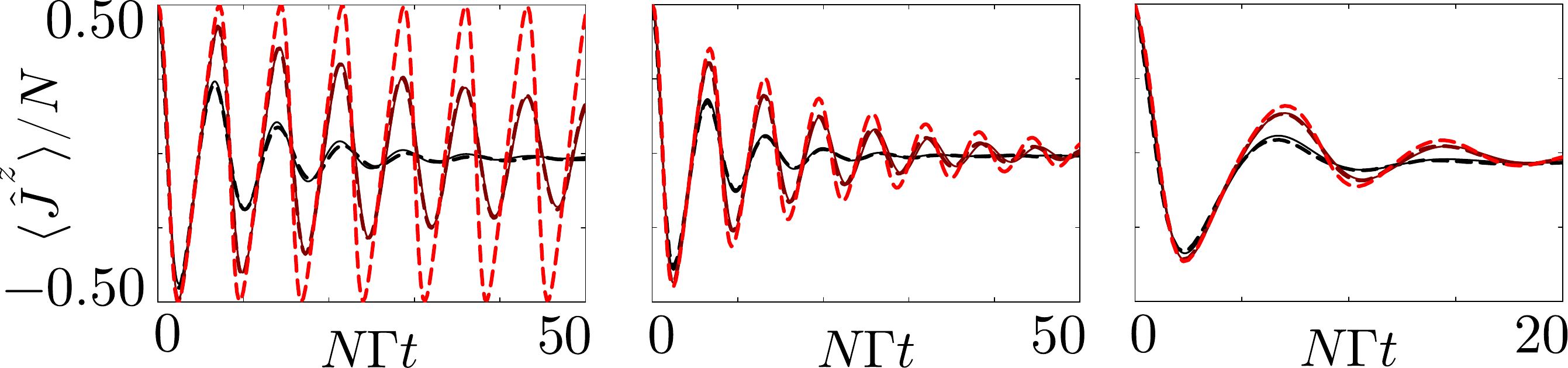}
    \caption{Evolution of the normalized inversion $\langle \hat{J}^{z}\rangle/N$ in a boundary time crystal without local dissipation (left) and with local dephasing (middle, $q=0$) or local decay (right, $q=-1$). Solid lines show exact results and dashed lines show TWA, for $N=10, 50, 10^4$ (dark to light), with $\Omega = N\Gamma$ and $\gamma = 0.2 N\Gamma$. Exact simulation is not feasible for $N=10^4$.}
    \label{supp-fig:time-crystal}
\end{figure}

While we focused on local pumping in Figs.~2--4 of the Main Text, the TWA readily applies to PI local decay and dephasing. We demonstrate this with the example of a boundary time crystal~\cite{supp-iemini18}. Here, an ensemble subject to coherent driving $\hat{H} = \Omega \hat{J}^{x}$ and collective decay with rate $\Gamma$ exhibits persistent oscillations of the inversion once $\Omega > N\Gamma/2$, approaching a limit cycle as $N$ becomes large. The impact of local dissipation on these oscillations has previously been demonstrated using exact methods for ensembles of tens of emitters in Ref.~\cite{supp-shammah2018}.\\
\noindent In Fig.~\ref{supp-fig:time-crystal}, we show the evolution of the normalized inversion for an ensemble in the time-crystal phase ($\Omega = N\Gamma$) subject to local dephasing or decay with rate $\gamma = 0.2 N\Gamma$. The TWA captures the damping of the oscillations, closely matching exact numerical results for $N=10$ and $50$. Using the TWA, we readily simulate a large system with $N = 10^4$, where the damping of the oscillations persists. These benchmarks demonstrate that the TWA can quantitatively describe the interplay of coherent Hamiltonian evolution with local dephasing and decay.

\subsection{Purely collective dynamics in the Schwinger-boson phase-space representation}
\label{supp-sec:collective-schwinger}

\begin{figure}[h!]
    \centering
    \includegraphics[width=0.33\linewidth]{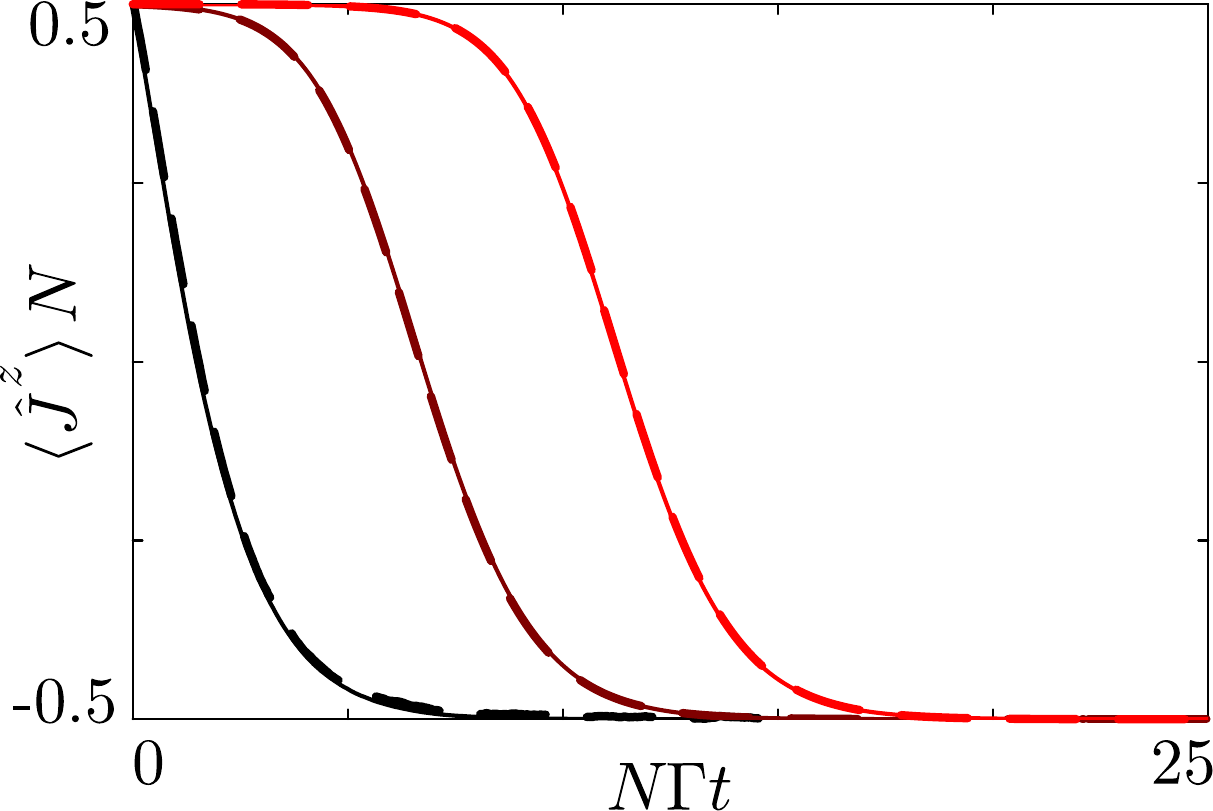}
    \caption{
Predictions of normalized inversion $\langle \hat{J}^{z}\rangle/N$ for superradiant decay with collective decay rate $\Gamma$. Exact solutions (solid) and TWA (dashed) are shown for (left to right) $N=3, 500, 5 \times 10^4$. Initially, all emitters are in the excited state.}
    \label{supp-fig:collective}
\end{figure}

\begin{figure}[h!]
    \centering
    \includegraphics[width=0.65\linewidth]{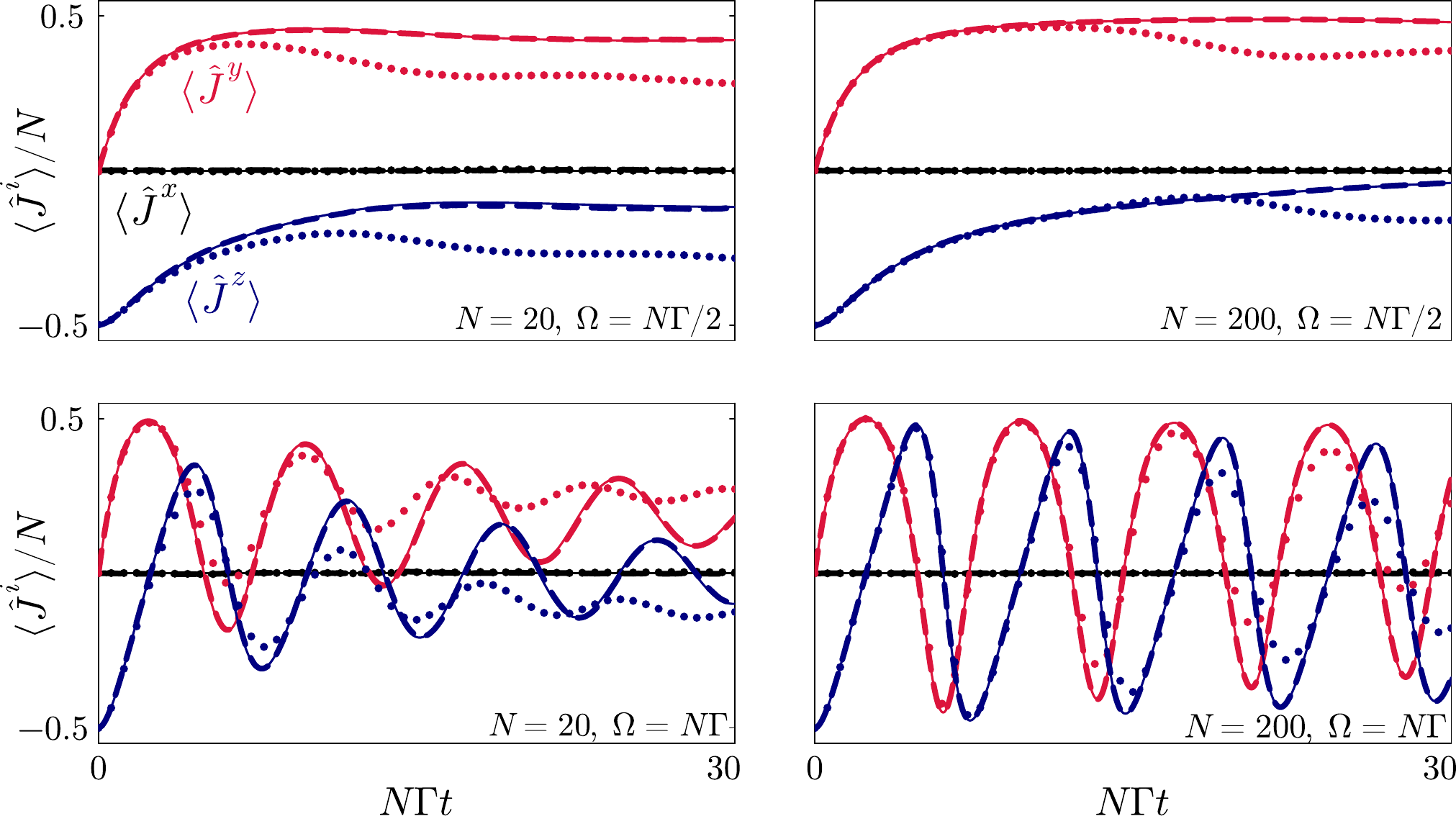}

    \caption{Evolution of the normalized collective spin observables $\langle \hat{J}^{i}\rangle/N$ $(i = x,y,z)$ in a boundary time crystal without local dissipation. Exact results are shown as solid lines, predictions of \cite{supp-huber21} as dots, and predictions of the present TWA as dashed lines.}
    \label{supp-fig:schwinger}
\end{figure}

In the Main Text, we introduced the TWA for PI ensembles with local dissipation. Here, we briefly illustrate application of the TWA and its relation to existing phase-space approaches in the example of purely collective dynamics. We first apply the TWA to superradiant decay (Fig.~\ref{supp-fig:collective}). The predictions agree well with exact simulations even for $N=3$ emitters, with agreement improving as $N$ increases. \\
\indent When collective dissipation is present, an approach in the phase space of Schwinger bosons has been applied to describe dynamics~\cite{supp-huber21}. Alongside the Wigner function, the dynamics of Schwinger bosons can be readily described in phase space using the Husimi-$Q$ and Glauber--Sudarshan $P$ phase-space representations, which correspond to anti-normal and normal ordering of bosonic operator products, respectively. \\
\indent In the $Q$ representation, Hamiltonian spin–spin interactions generate contributions to diffusion.  For squeezing, these contributions can make the diffusion matrix indefinite, precluding direct Monte Carlo sampling. However, discarding these terms to restore positivity compromises accuracy in predictions of squeezing dynamics. In the $P$ distribution, on the other hand, squeezed states cannot be represented by a nonnegative distribution and therefore cannot be numerically captured by direct stochastic sampling. The positive-$P$ representation~\cite{supp-drummond80} can overcome these problems and capture quantum dynamics by treating the bosonic amplitudes and their conjugates as independent complex variables on a doubled phase space, although trajectory instabilities can restrict the useful simulation time~\cite{supp-chuchurka24}. The TWA considered here avoids these difficulties by perturbatively incorporating leading quantum fluctuations~\cite{supp-polkovnikov10} into nonlinear stochastic dynamics on the collective spin’s classical phase space. \\
\indent  The TWA obtained for Schwinger bosons using the semiclassical expansion of \cite{supp-dubois2021} agrees with our formulation for collective dynamics, upon neglecting relative $O(\mathcal J^{-2})$ corrections and expressing the equations in collective-spin variables. However, a direct expansion of the Lindbladian in phase space and truncation at second order of derivatives does not generally produce positive diffusion for the Wigner function evolving under a nonlinear Lindbladian~\cite{supp-carmichael2008}. In \cite{supp-huber21}, additional diffusion terms were removed to mitigate this effect in collective dynamics, neglecting terms relevant in the semiclassical expansion. Initial Schwinger-boson number states were also approximated by coherent states, introducing additional fluctuations in the spin length. To illustrate their effect, in Fig.~\ref{supp-fig:schwinger} we compare predictions of the present TWA to the approach in \cite{supp-huber21} for the model of a boundary time crystal, studied in Fig.~\ref{supp-fig:time-crystal}, in the absence of local dissipation. For $N = 20$ and $200$ emitters, predictions of \cite{supp-huber21} deviate from the exact solution over longer times, with the discrepancy visible for $N=200$ at the phase transition to the boundary time crystal $(\Omega = N\Gamma/2)$. By contrast, the present TWA closely tracks the exact expectations throughout the displayed evolution. We note that discrepancies remain even when using the improved initial sampling of Ref.~\cite{supp-altland09}, as relevant corrections remain missed in the semiclassical equations of motion. Additionally using the TWA obtained through the semiclassical expansion of Schwinger bosons~\cite{supp-dubois2021} recovers close agreement with the present TWA.
\subsection{Comparison with cumulant expansion in the dynamics of a driven-dissipative system}
\begin{figure}[h!]
    \centering
    \includegraphics[width=0.75\linewidth]{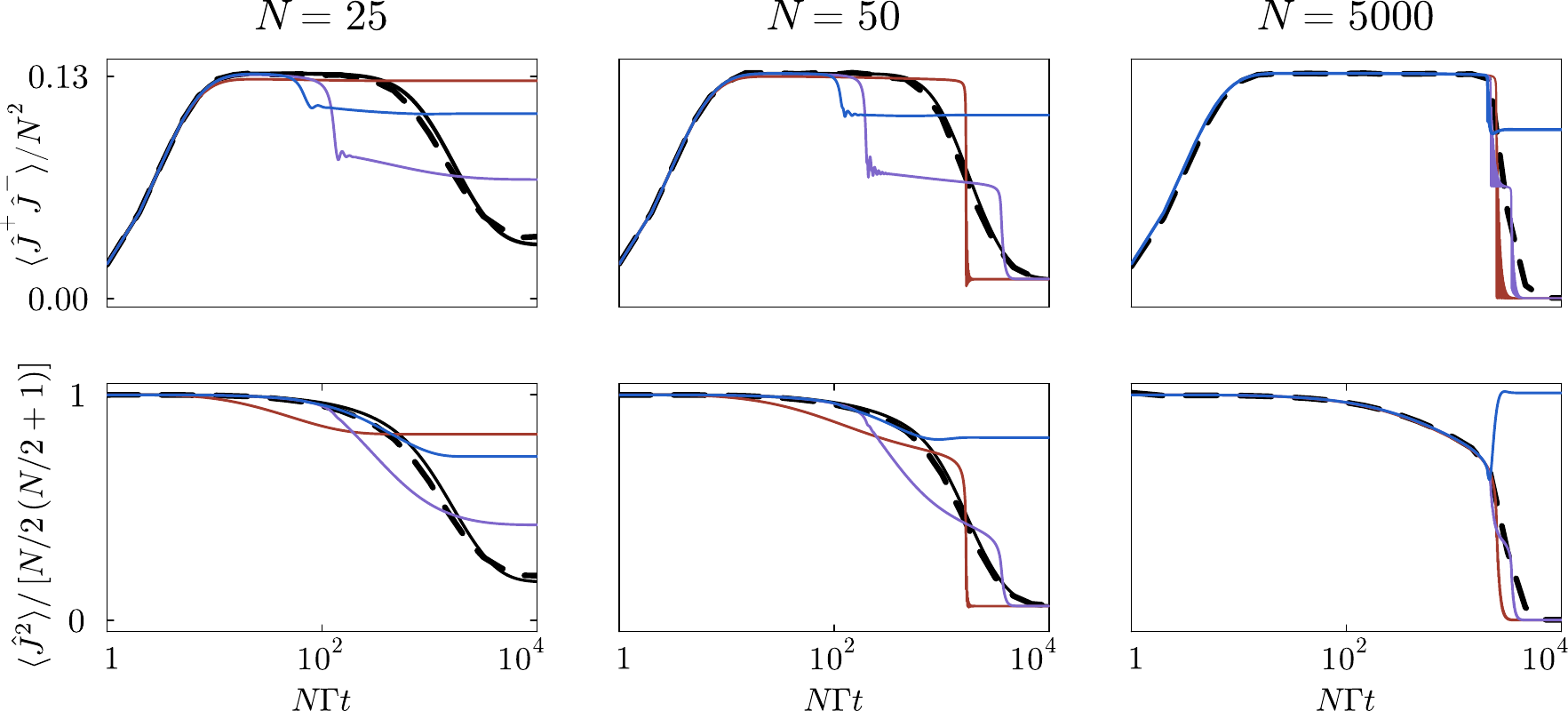}
    \caption{Evolution, on a log-time scale, of the normalized emission rate $\langle \hat{J}^{+}\hat{J}^{-} \rangle/N^2$ (top) and squared spin length $\langle\hat J^2\rangle/[N/2(N/2+1)]$ (bottom) in a driven-dissipative ensemble of $N = 25, 50, 5000$ emitters. Exact results are shown as solid black curves where available, and TWA predictions are given as dashed black curves. Predictions under first-, second-, and third-order cumulant closures are shown as red, lavender, and navy curves, respectively. Exact simulation is not feasible for $N=5000$.}
    \label{supp-fig:cumulant}
\end{figure}

\indent 
In the Main Text, we showed how the TWA can provide quantitative improvements in accuracy when predicting steady-state expectations compared to the second-order cumulant expansion. In dynamics, non-Gaussian correlations can become especially pronounced and lead to qualitative errors in cumulant expansion predictions~\cite{supp-stitely23,supp-fasser25}. In Fig.~\ref{supp-fig:cumulant}, we present evolution of a driven-dissipative model in which coherent driving $\hat{H} = \Omega \hat{J}^{x}$ competes with collective and local decay, with rates $\Gamma$ and $\gamma$, respectively~\cite{supp-Bonifacio1976,supp-leppenen2024}. Choosing $\Omega/(N\Gamma)=0.365$, $\gamma/(N\Gamma)=0.004$, and an initial ground state, we observe a rapid increase in collective emission on the timescale $(N\Gamma)^{-1}$, followed by a crossover on the longer timescale $\gamma^{-1}$ marked by reductions in collective emission and squared spin length. First-order (mean-field), second-, and third-order cumulant closures of single-emitter operators obtained using Ref.~\cite{supp-plankensteiner22} quantitatively track the fast dynamics, but develop abrupt changes and oscillations in collective emission that are absent in exact dynamics for $N=25$ and $50$ during the crossover. These features persist with smaller solver time steps and tighter tolerances. The TWA here tracks the exact expectations over the full evolution for $N=25$ and $50$. In this model, $\langle\hat J^2\rangle$ remains much greater than unity throughout evolution, consistent with the observed improvement of the TWA with increasing emitter number. While the cumulant expansion predictions improve with increasing $N$ at earlier times, oscillations persist through the crossover up to $N=5000$, where the displayed sequence shows no consistent convergence with increasing closure order up to third order. \\
\indent This example illustrates that, even at large $N$, fixed low-order cumulant closures need not yield converged predictions for nonequilibrium dynamics that remain accessible within the TWA. Because the TWA does not impose such a closure, it also provides an independent, scalable comparison with cumulant-expansion predictions at emitter numbers beyond the reach of exact simulation.

\subsection{Strongly nonlinear quantum dynamics in the TWA}
\begin{figure}[h!]
    \centering
    \includegraphics[width=0.75\linewidth]{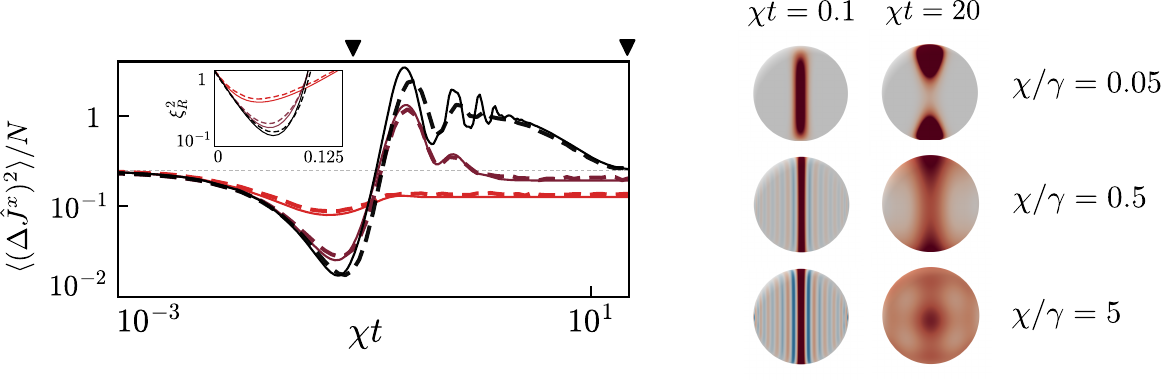}

    \caption{Log-time evolution of the normalized variance $\langle (\Delta \hat{J}^{x})^{2} \rangle/N$ in an ensemble of $N = 25$ emitters, for normalized twisting strength  $\chi/\gamma = 0.05,0.5,5$ (red to black). Exact results are shown as solid curves  and TWA predictions are given as dashed curves. The horizontal dashed line at $1/4$ corresponds to a $z$-polarized coherent state. Inset shows short-time evolution of the Wineland squeezing parameter, $\xi_{\mathrm{R}}^{2} = N\langle (\Delta \hat{J}_{\mathrm{min}})^2 \rangle / |\langle\hat{\mathbf J}\rangle|^2$,  for variance $\langle (\Delta \hat{J}_{\mathrm{min}})^2\rangle$ minimized over quadratures orthogonal to $\langle\hat{\mathbf J}\rangle$. Right panel: exact Wigner functions, viewed from the south pole, in the most-populated total-spin sector at times $\chi t = 0.1$ and $20$, indicated by triangular markers in the evolution plot. At $\chi t=0.1$, $\mathcal{J}=N/2$ for all rows; at $\chi t \simeq 20$, $\mathcal{J}=10.5,4.5,3.5$ from top to bottom. Red and blue denote positive and negative regions. The Wigner function is independent of $\psi$, which is not shown.}
    \label{supp-fig:squeeze}
\end{figure}

In Fig.~2(b) of the Main Text, we identified greater discrepancies between the TWA and exact evolution for decreasing spin length. In addition to macroscopic spin length, the validity of the semiclassical expansion requires the Wigner function to remain smooth on the effective Planck scale~\cite{supp-habib1998}, so that higher-order terms in the semiclassical expansion can be safely neglected. We briefly illustrate this case in the current TWA using the nonlinear two-axis counter-twisting Hamiltonian $\hat{H} = \frac{\chi}{2i}(\hat{J}^{+2} - \hat{J}^{-2})$ in the presence of uniform local decay. In Fig.~\ref{supp-fig:squeeze}, we plot the evolution of squeezing under competition with uniform local decay for $\chi/\gamma = 0.05,  0.5,  5$, beginning from the fully $z$-polarized ground state. The TWA captures the initial squeezing, followed by relaxation under local dissipation. At the onset of visible deviations ($\chi t \simeq 0.1$) for weak decay, exact simulations give $\langle \hat{J}^2 \rangle /[N/2(N/2+1)] \gtrsim 0.9$ while the total-spin variable $\mathcal J$ remains macroscopic in TWA trajectories, showing that the discrepancies begin before substantial reduction of the spin length. To examine the larger quantitative discrepancy between the TWA and the exact solution for weaker local decay, we plot the exact Wigner function $W(\phi,\theta,\mathcal{J} = N/2)$ in the maximally populated sector at $\chi t = 0.1$. Larger quantitative discrepancies are found when the Wigner function exhibits more rapid oscillations, with quantum interference accompanied by appreciable Wigner negativity. At late times $(\chi t = 20)$, diffusion of the Wigner function under local decay reduces fast variation, consistent with the recovered agreement between the TWA and exact dynamics. In the cases considered, close agreement is maintained throughout when local decay is comparable to or faster than twisting.  

\subsection{Conditioned field trajectories in the TWA}

\subsubsection{Correspondence between semiclassical field trajectories and heterodyne measurement records}

In the Main Text, we presented individual trajectories of semiclassical signals for fields exiting the ends of the waveguide. For completeness, we here provide a mathematical argument showing how trajectories generated using the semiclassical SDE in Eq.~(6) of the Main Text can be related to heterodyne measurement records within the TWA. For simplicity, we consider a single ensemble, as the result for several PI ensembles readily follows. We assume that an ideal heterodyne measurement is performed on an output port corresponding to a physical jump channel $\hat L_\kappa$ that is invariant under permutation of emitters. For heterodyne measurements of the output-channel jump operator $\hat{c} = \sqrt{r_\kappa}\hat L_\kappa$, the density matrix $\hat\rho_c$ conditioned on the observed complex record $Z$ evolves according to the It\^o stochastic differential equation,
\begin{align}
    d\hat\rho_c
    &=
    \left[-i\left[\hat{H},\hat{\rho}_c\right]
    + \sum_{q}{\frac{\Gamma^{q}}{2}}\mathcal{D}\left[\hat{L}^{q}\right]\hat{\rho}_c
    + \sum_{q,n}{\frac{\gamma^{q}}{2}}\mathcal{D}\left[\hat{\sigma}_{n}^{q}\right]\hat{\rho}_c\right]d t
    + \left(\hat c-\langle\hat c\rangle_c\right)\hat\rho_c\,d\nu^*
    + \hat\rho_c\left(\hat c^\dagger-\langle\hat c^\dagger\rangle_c\right)d\nu,
    \label{supp-eq:compact-sme}\\
    d Z
    &=
    \langle \hat c \rangle_c\,d t
    +
    d\nu .
    \label{supp-eq:compact-quantum-record1}
\end{align}
Here the complex Wiener increment $d\nu$ is the innovation under measurement, with $\langle d\nu\rangle= \langle d\nu d\nu \rangle = 0$ and
$\langle d\nu d\nu^* \rangle=dt$, such that averaging Eq.~\eqref{supp-eq:compact-sme} over $d\nu$ recovers the
unconditional master equation, Eq.~(1) of the Main Text.

For the PI initial states and dynamics considered here, Eq.~\eqref{supp-eq:compact-sme} preserves permutation invariance and may be interpreted in the angular-momentum representation.
There, tracing against the Stratonovich--Weyl kernel
$\hat\Delta(\bm{\mathcal Z})$ and applying the semiclassical expansion, Eq.~(5) of the Main Text, yields an It\^o Fokker--Planck equation for the evolution of the conditioned Wigner function $W_c = \mathrm{Tr}[\hat{\Delta}\hat{\rho}_c]$ in the TWA, written in the canonical phase-space coordinates $\bm{\mathcal Z}'$,
\begin{align}
    d W_c
    &=
    \left[
        -\sum_{i}\frac{\partial}{\partial \mathcal{Z}_i'}\!\big[u_i(\bm{\mathcal{Z}}')\,W_c\big]
        +\frac{1}{2}\sum_{ij}\frac{\partial^2}{\partial \mathcal{Z}_i'\,\partial \mathcal{Z}_j'}\!\big[D_{ij}(\bm{\mathcal{Z}}')\,W_c\big]
    \right]d t
    \nonumber\\
    &\quad+
    \left[
        \left(c(\bm{\mathcal Z}')-\bar c_c\right)W_c
        -
        \sum_i
        \frac{\partial}{\partial\mathcal Z_i'}
        \big[\mathcal C_i(\bm{\mathcal Z}')W_c\big]
    \right]d\nu^*
    +
    \left[
        \left(c(\bm{\mathcal Z}')^*-\bar c_c^*\right)W_c
        -
        \sum_i
        \frac{\partial}{\partial\mathcal Z_i'}
        \big[\mathcal C_i(\bm{\mathcal Z}')^*W_c\big]
    \right]d\nu,
    \label{supp-eq:compact-conditional-fpe}
    \\
    d Z
    &=
    \bar{c}_c \,d t
    +
    d\nu,
    \label{supp-eq:compact-quantum-record}
\end{align}
for the Weyl symbol $c(\bm{\mathcal Z}')=\sqrt{r_\kappa}\mathcal L_\kappa(\bm{\mathcal Z}')$ and the expectation $\bar{c}_c$ taken with respect to $W_c$. In the TWA, correlations between the observation noise and the conditional phase-space evolution are encoded by the semiclassical correction,
\begin{equation}
    \mathcal C_i(\bm{\mathcal Z}')= \frac{i}{2}\left\{\mathcal Z_i', c(\bm{\mathcal Z}')\right\}.
    \label{supp-eq:measurement-C}
\end{equation}
Eq.~\eqref{supp-eq:compact-conditional-fpe} has the form of a Kushner--Stratonovich filtering equation~\cite{supp-Altafini2012} describing the evolution of the Wigner function $W_c$ conditioned on the measurement record $Z$. The same evolution is obtained by conditioning on trajectories $Z$ generated by the explicit Stratonovich stochastic differential equations~\cite{supp-Xiong2008},
\begin{align}
    d\bm{\mathcal{Z}}'
    &= \{ \bm{\mathcal{Z}}',\mathcal{H}\}dt
    + \mathrm{Re}\!\left[
        \left\{\bm{\mathcal{Z}}',c\right\}\circ
        \Bigl(d W_\kappa+i c^*dt\Bigr)
        +\sum_{k\ne\kappa}\left\{\bm{\mathcal{Z}}',\sqrt{r_k}\mathcal L_k\right\}\circ
        \Bigl(dW_k+i\sqrt{r_k}\mathcal L_k^*dt\Bigr)
      \right],
      \label{supp-eq:SDE-differential1}\\
    dZ
    &=c(\bm{\mathcal Z}') dt+\frac{i}{2}\left(dW_\kappa^* - dB_\kappa\right),
      \label{supp-eq:SDE-differential2}
\end{align}
for complex Wiener increments with $\langle d W_k d W_{k'}^*\rangle = 2dt \delta_{kk'}$ and
$\langle d W_k d W_{k'}\rangle=0$. The increments $dB_\kappa$ represent additional vacuum noise under heterodyne measurement. They are independent of all $dW_k$ and likewise satisfy $\langle d B_\kappa d B_\kappa^*\rangle = 2dt$, $\langle d B_\kappa^2 \rangle = 0$. In Eq.~\eqref{supp-eq:SDE-differential2}, $Z$ is then calculated along a single realized trajectory $\bm{\mathcal{Z}}'$, with the observation noise correlated with the TWA trajectory through the shared increment $dW_\kappa$. 

Crucially,  Eq.~\eqref{supp-eq:SDE-differential1} gives the semiclassical phase-space trajectories of Eq.~(6) of the Main Text in differential form and canonical coordinates. Conditioning an ensemble of these trajectories on the associated records $Z$ yields the distribution $W_c$ governed by Eq.~\eqref{supp-eq:compact-conditional-fpe}. The joint SDEs in Eqs.~\eqref{supp-eq:SDE-differential1} and \eqref{supp-eq:SDE-differential2} therefore generate records with the same path distribution as the record process associated with Eqs.~\eqref{supp-eq:compact-conditional-fpe} and \eqref{supp-eq:compact-quantum-record}. This equivalence is exact within the semiclassical model. When the TWA accurately approximates the exact conditioned quantum dynamics, these generated records can also be interpreted as approximate experimental heterodyne records. For a joint measurement of the forward and backward waveguide output ports, relating to Fig.~4 of the Main Text, $\nu = \mathrm{F,B}$, the complex heterodyne records are similarly generated by the relation
\begin{equation}
dZ_\nu = c_\nu(\bm{\mathcal Z}')dt + \frac{i}{2}\left(dW_\nu^*-dB_\nu\right), 
\qquad
c_\nu=\sqrt{\Gamma/2}\,\mathcal L_\nu,
\end{equation}
where the noise increments for the two ports are independent. In the Main Text, we display the corresponding forward and backward semiclassical signal trajectories and omit the accompanying white-noise contribution to more simply present the qualitative dynamics.

\subsubsection{Numerical verification of the semiclassical correspondence for conditioned dynamics}

\begin{figure}[h!]
    \centering
    \includegraphics[width=0.65\linewidth]{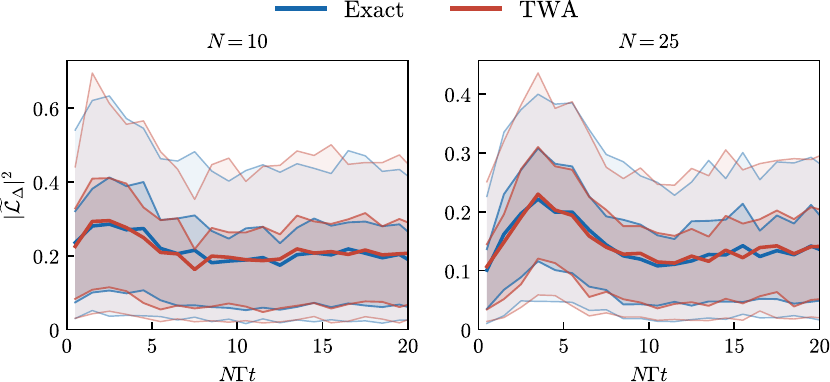}
    \caption{Distribution of the normalized binned heterodyne-record power $|\widetilde{\mathcal L}_{\Delta}|^2$ for $N=10$ and $25$, using the parameters of Fig.~2(a) in the Main Text. Exact conditioned dynamics and TWA predictions are shown in blue and red, respectively. Thick curves show the mean, while dark and light shaded regions span the central $50\%$ and $80\%$ of trajectories, respectively. The bin width satisfies $N\Gamma\Delta t=1$. We use $256$ trajectories in each case.}
    \label{supp-fig:heterodyne-validation}
\end{figure}

As a finite-size numerical check in the setting of PI ensembles, we consider the same single-ensemble model and parameter scaling as in Fig.~2(a) of the Main Text, with a fully inverted initial state, $\hat H=0$, collective decay at rate $\Gamma$, and unmonitored local pumping at $\gamma=0.25N\Gamma$. We assume ideal heterodyne monitoring of the collective-decay output channel $\hat c=\sqrt{\Gamma}\hat J^-$. Exact records are generated from the heterodyne stochastic master equation in the PI basis using Eqs.~\eqref{supp-eq:compact-quantum-record1} and \eqref{supp-eq:compact-sme}, while TWA records are generated from Eqs.~\eqref{supp-eq:SDE-differential1} and \eqref{supp-eq:SDE-differential2}.\\
\indent The heterodyne records contain white noise, so we define the normalized binned record $\widetilde{\mathcal L}_{\Delta}(t)
=(\sqrt{\Gamma}N\Delta t)^{-1}
\int_t^{t+\Delta t}dZ(s)$, with $N\Gamma\Delta t=1$. This quantity is a noisy semiclassical estimate corresponding to the normalized, bin-averaged conditional amplitude $\langle\hat J^-\rangle_c/N$. Figure~\ref{supp-fig:heterodyne-validation} presents the time-dependent distributions of powers obtained from these records, showing close visual agreement between the exact and TWA means and central quantile bands for $N=10$ and $25$. This comparison supports a TWA description of heterodyne measurement records. \\
\indent During superradiant emission in Fig.~4 of the Main Text, an intermediate bin width may be chosen over which the detected signal is large compared with shot noise while remaining short compared with the characteristic evolution time. The binned record then approximates the instantaneous semiclassical field without appreciably averaging its dynamics, and we therefore present the field signal to qualitatively describe the trajectory-level directional dynamics.

\subsection{Application of the TWA to the central-spin model}

\begin{figure}[h!]
    \centering
    \includegraphics[width=0.33\linewidth]{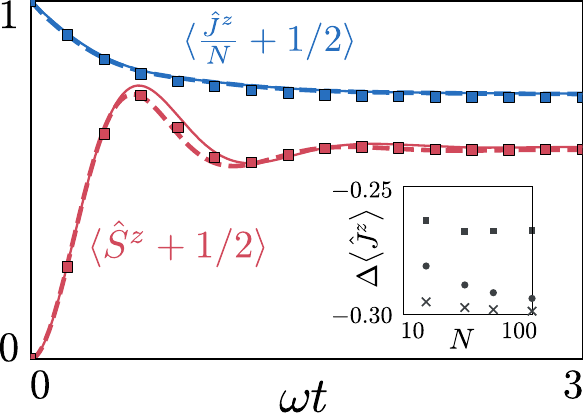}
    \caption{Evolution of the central-emitter excitation $\langle\hat{S}^{z} + 1/2\rangle$ (red) and the satellite excitation fraction $\langle\hat{J}^{z}/N + 1/2\rangle$ (blue) for $N=30$. Thin solid curves show exact results, thick dashed curves show the present TWA, and squares show the site-resolved TWA of Ref.~\cite{supp-hosseinabadi25}. The inset shows the change $\Delta\langle\hat{J}^{z}\rangle$ in the satellite inversion relative to the noninteracting case $g=0$, versus $N$ at $\omega t=3$. Crosses, circles, and squares denote the exact, present, and site-resolved TWA results, respectively. Parameters are $\omega=\epsilon=\kappa=1$, $g=3$, $\gamma_\downarrow=1/2$, and $\gamma_\uparrow=3/2$.}
    \label{supp-fig:central-spin}
\end{figure}

In the Appendix, we discussed how a semiclassical description of dynamics can be established for hybrid systems partitioned into homogeneous ensembles and individual emitters. Here, we validate its application in the central-spin model,
\begin{equation}
    \hat{H}
    =\omega\hat{S}^{z}
    +\epsilon\hat{J}^{z}
    +\frac{g}{\sqrt{N}}
    \left(\hat{S}^{+}\hat{J}^{-}
    +\hat{S}^{-}\hat{J}^{+}\right),
    \label{supp-eq:central-spin-H}
\end{equation}
where $\hat{S}^{\alpha}$ denotes the spin operators of the central emitter and $\hat{J}^{\alpha}$ are the collective-spin operators of the $N$ satellite emitters. The central emitter undergoes decay at rate $\kappa$, while each satellite emitter undergoes independent decay and pumping at rates $\gamma_\downarrow$ and $\gamma_\uparrow$, respectively. The central emitter is initially in its ground state and the satellite ensemble is fully excited.
Following the Appendix, we identify the central spin as an ensemble of size one and collect the satellite spins into a single PI ensemble of size $N$. For the initial ground state of the central emitter, we use the discrete sampling prescription~\cite{supp-hosseinabadi25} (see ``\textit{Sampling the initial state Wigner function}'') for the phase-space variables $\mathbf{S}$,
\begin{equation}
    \mathbf{S}=(S_x,S_y,-1/2),
    \qquad S_x,S_y\in\{-1/2,1/2\},
    \label{supp-eq:central-spin-sampling}
\end{equation}
where $S_x$ and $S_y$ are sampled independently with equal probabilities. The initial state of the collective satellite spin is sampled using the Gaussian approximation for the corresponding Wigner function. We evolve Eq.~(6) of the Main Text, using the standard Poisson bracket relations for the single spin (see \textit{``Recovering the TWA for individual emitters''}), and the Poisson bracket of the collective satellite spin is understood as in the Main Text. \\
\indent In Fig.~\ref{supp-fig:central-spin}, the present TWA closely reproduces the excitation dynamics of the central and satellite spins for $N=30$. For reference, we also show the prediction of the site-resolved TWA, for which every emitter is assigned an independent intrinsic two-dimensional spin phase space~\cite{supp-hosseinabadi25}. The present construction uses two intrinsic phase-space variables for the central emitter and four for the satellite ensemble for a total of six phase-space variables independent of $N$, compared with $2(N+1)$ in the site-resolved formulation. The present TWA also captures the order-unity change $\Delta\langle\hat J^z\rangle$ in the satellite inversion, induced by the central spin, relative to the noninteracting case $g=0$. The discrepancy from the exact result also decreases with increasing $N$, with close quantitative agreement at $N=100$. This example shows that the present TWA can remain accurate for an individual emitter coupled to a large PI ensemble.

\begingroup
\makeatletter
\let\thebibliography\NAT@thebibliography
\let\endthebibliography\endNAT@thebibliography
\let\NAT@bibsetnum\NATx@bibsetnum
\makeatother
\input{supplement.bbl}
\endgroup

\end{document}